\documentclass{report}

\usepackage{amsfonts,amsmath,amssymb}
\usepackage[dvips]{graphics}
\usepackage{epsfig}
\usepackage{amscd}
\usepackage{verbatim}
\usepackage{euscript}
\usepackage{color}
\newcommand{\Rl}{\mathbb{R}}
\newcommand{\Nl}{\mathbb{N}}
\newcommand{\Ir}{\mathbb{Z}}
\newcommand{\Cx}{\mathbb{C}}
\newcommand{\A}{\mathcal{A}}
\newcommand{\HH}{\mathcal{H}}

\newcommand{\s}{\sigma}
\newcommand{\C}{\mathcal{C}}
\newtheorem{theorem}{Theorem}[section]
          
\newtheorem{lemma}[theorem]{Lemma}              
              
\newtheorem{proposition}[theorem]{Proposition}

\newtheorem{definition}[theorem]{Definition}
\newtheorem{conjecture}[theorem]{Conjecture}
\newcommand{\rem}[1]{{\bf Remark:}}

\newcommand{\Section}[1]{\setcounter{equation}{0}\section{#1}}

\newenvironment{proof}{\noindent {\bf Proof: }}{\QED\medskip}
\def\QED{{\hspace*{\fill}{\vrule height 1ex width 1ex }\quad} 
    \vskip 0pt plus20pt}

\def\idty{{\mathchoice {\rm 1\mskip-4mu l} {\rm 1\mskip-4mu l} % 
{\rm 1\mskip-4.5mu l} {\rm 1\mskip-5mu l}}}
\newcommand{\be}{\begin{equation}}
\newcommand{\ee}{\end{equation}}
\newcommand{\benn}{\begin{equation*}}
\newcommand{\eenn}{\end{equation*}}
\newcommand{\bea}{\begin{eqnarray}}
\newcommand{\eea}{\end{eqnarray}}
\newcommand{\beann}{\begin{eqnarray*}}
\newcommand{\eeann}{\end{eqnarray*}}

\newcommand{\ket}[1]{\vert{#1}\rangle}
\newcommand{\bra}[1]{\langle{#1}\vert}

\newcommand{\unity}{{1\hskip -3pt \rm{I}}}

\newcommand{\ip}[2]{\langle{#1}|{#2}\rangle}

\newcommand{\OP}{\operatorname{P}}
\newcommand{\obs}{\mathcal{A}}

\newcommand{\nn}[2]{\langle #1,#2 \rangle}
\newcommand{\tr}{\mathrm{Tr}}

\begin{document}
\pagenumbering{roman}
\begin{center}
{\Large \bf Low Temperature Results for the Heisenberg XXZ and XY Models}
\vskip 0.2in
\centerline{By}
\vskip 0.1in
\centerline{\large  Justin Abbott}
\centerline{\large B.A. (University of California, Davis) 1996}

\vskip 0.3in
\centerline{\large  DISSERTATION}
\vskip 0.1in
\centerline{\large Submitted in partial satisfaction of the requirements for the
degree of}
\vskip 0.1in
\centerline{\large  DOCTOR OF PHILOSOPHY}
\vskip 0.1in
\centerline{\large in}
\vskip 0.1in
\centerline{\large MATHEMATICS}
\vskip 0.2in
\centerline{\large in the}
\vskip 0.1in
\centerline{\large  OFFICE OF GRADUATE STUDIES}
\vskip 0.1in
\centerline{\large of the}
\vskip 0.1in
\centerline{\large  UNIVERSITY OF CALIFORNIA}
\vskip 0.1in
\centerline{\large  DAVIS}
\end{center}
\vskip 0.1in
{\large Approved:}
\begin{center}
\centerline{\underbar{\hskip 2.5in}}
\vskip 0.15in
\centerline{\underbar{\hskip 2.5in}}
\vskip 0.15in
\centerline{\underbar{\hskip 2.5in}}
\vskip 0.2in
\centerline{\large Committee in Charge}
\vskip 0.2in
\centerline{\large 2003}
\end{center}

\newpage
\large
\tableofcontents

%%%%%%%%%%%%%%%%%%%%%%%%%%%%%%%%%%%%%%%%%%%%%%%%%%%%%%%%%%%%%%%%%%
%%%%%%%%%%%%%%%%%%%%%%%%%%%%%%%%%%%%%%%%%%%%%%%%%%%%%%%%%%%%%%%%%%

\newpage
\large
{\Large \bf ACKNOWLEDGEMENTS} \\

I have been very lucky to be a part of the UC Davis math department for the last
six years.  My advisor, Bruno Nachtergaele, has been an amazingly patient and
inspirational teacher and I thank him for everything.  Wolfgang Spitzer and
Daniel Ueltschi have been no less than second advisors to me.  I wish them well
in their future careers.  I would also like to thank my mother Joanna, my
father  Daniel, and my ``little'' brother Morgan, for all their support
throughout each stage of my degree. Finally, I must acknowledge that without 
my wife Jennifer, and my children Harper and Hudson, I would never have been able to
find the will to complete my Ph.D.  Their love and support have been more
important to me than they can possibly know.

%%%%%%%%%%%%%%%%%%%%%%%%%%%%%%%%%%%%%%%%%%%%%%%%%%%%%%%%%%%%%%%%%%%%%%%%%%%%%%%%%%%%%%%%%%%%%%%%%%%%%%%%%%%%%%%%%%%%%%%%%%%%%%%%%%%%%%%%%%%%%%%%%%%%%%%%%%

\newpage 
\begin{center}
\underline{\bf \Large Abstract} 
\end{center}

\medskip

\large
This thesis contains two results for the low temperature behavior of quantum
spin systems.
First, we present a lower bound for the spin-1 XXZ chain in finite volumes in terms of 
the gap of the two-site Hamiltonian.  The estimate is derived by a method 
developed by Nachtergaele in \cite{Na2} called the Martingale Method.  Our bound 
relies on an assumption which we have, as yet, been unable to verify 
analytically in all cases.  We present numerical evidence that strongly 
indicates our assumption is valid.  

The second result is a proof that the 
spin-1/2 $d$-dimensional XY model in the presence of an external magnetic field 
does not undergo a phase transition at low temperature, provided that the 
strength of the field is great enough.  Using a contour expansion inspired by 
Kennedy in \cite{Ken}, we show that the weights of contours satisfy a condition 
of Koteck{\'y} and Preiss \cite{KP} which allows us to express the free energy 
of the system as a cluster expansion.  As part of the setup we give a simple 
proof that the all-spin-up state is the unique ground state when the 
external magnetic field has strength at least $2d$.

%%%%%%%%%%%%%%%%%%%%%%%%%%%%%%%%%%%%%%%%%%%%%%%%%%%%%%%%%%%%%%%%

\newpage
\pagestyle{myheadings} 
\pagenumbering{arabic}
\markright{  \rm \normalsize CHAPTER 1. \hspace{0.5cm}
 Introduction }
\large 
\chapter{Introduction}
\label{chapter: introduction}
\thispagestyle{myheadings}

%%%%%%%%%%%%%%%%%%%%%%%%%%%%%%%%%%%%%%%%%%%%%%%%%%%%%%%%%%%%%%%%%%%%%%%%%%%%%%%%
%%%%%%%%%%%%%% CHAPTER: INTRODUCTION %%%%%%%%%%%%%%%%%%%%%%%%%%%%%%%%%%%%%%%%%%%
%%%%%%%%%%%%%%%%%%%%%%%%%%%%%%%%%%%%%%%%%%%%%%%%%%%%%%%%%%%%%%%%%%%%%%%%%%%%%%%%

This thesis contains two results in the statistical mechanics of quantum spin
systems.  The first is an explicit lower bound bound for the spectral gap of the 
one dimensional, spin-1, XXZ ferromagnet.  The estimate is obtained using a 
method developed by Nachtergaele in \cite{Na2}.  The second result is a proof on 
the absence of a phase transition for the Heisenberg XY  model in an external 
magnetic field.  In general the XXZ and XY models have very different behavior, 
ironically however, the techniques used in Chapter \ref{chapter: xy} were 
developed by Kennedy in \cite{Ken} to treat the low  temperature XXZ model.  

 The XXZ Hamiltonian on a domain $\Lambda\subset\Ir^d$ is given by
 \be
 H^{XXZ}_\Lambda =
 -\sum_{\stackrel{\{x,y\}\subset\Lambda}{|x-y|=1}}\frac{1}{\Delta}S^1_xS^1_y+\frac{1}{\Delta}S^2_xS^2_y+S^3_xS^3_y
 \ee
 where the operators $S^\alpha_x$ are the spin operators in the direction $\mathbf{e_\alpha}$ 
 at the site $x\in\Lambda$, $\Delta>1$, and 
 $|\cdot|$ is the standard $\ell^1$-norm on $\Ir^d$.  Ferromagnets are of 
 general physical interest, and are
 not uncommon in nature.  However there are other reasons to study ferromagnets
 as opposed to antiferromagnets or ferrimagnets.  The first, and perhaps 
 least serious, is that it tends to be very difficult to extract information from
 antiferromagnetic models. There are good results on antiferromagnets though. 
 For example in \cite{DLS} Dyson, Lieb, and Simon used reflection positivity
 to prove the existence of long range order in dimension two or greater. However,
 the XXZ model is not, in general, reflection positive \cite{Speer}.
 
 For the ferromagnet much more is known.    In \cite{ASW} Alcaraz, Salinas, and
 Wreszinski gave a formula for the ground states of the XXZ chain for any spin. 
 Koma and Nachtergaele showed that in the thermodynamic limit, the spin-1/2 
 XXZ model has
 a spectral gap above the ground state, and were able to calculate an exact
 expression for the gap \cite{KN1}.  Later, Matsui used the existence of the gap
 to prove that for any anisotropy $\Delta>1$, the list of ground states
 given in \cite{ASW} is indeed a complete list.  Matsui's result was then
 improved upon by Koma and Nachtergaele in \cite{KN3} where they used the
 intrinsic quantum group symmetry of the XXZ chain to extend Matsui's result to
 the case $\Delta=1$.  
 
 The existence of a quantum group symmetry of the finite
 volume spin-1/2 chain, initially noted by Pasquier and Saleur in \cite{PS}, is
 still another reason for studying the XXZ model.  Physcists generally associate
 any symmetry with physical meaning, while mathematicians tend to view it as a
 special but interesting situation that is deserving of further investigation.  
 In \cite{KN1} Koma and Nachtergaele not only calculated the spectral gap of the
 spin-1/2 infinite volume chain, but proved a sort of ordering of energy levels 
 (\`{a} la Lieb and Mattis in \cite{LM}) for the finite chain.   Specifically
 they showed that not only is the ground state  of $H^{XXZ}_L$ the $L+1$
 dimensional irreducible representation of $SU_q(2)$, but any first excited
 state must lie in an $L-1$ dimensional irreducible representation.  
 The quantum group
 symmetry does have some limitations however.  It only exists in a restricted
 setting, i.e. for spin-1/2 chains in one dimension.  In this paper we are
 interested in the spin-1 chain, so other tools must be used.
 
 The low-energy spectrum of the XXZ chain is also fairly well understood.  This
 is relevant because the low-energy spectrum, and in particular the spectral
 gap, have direct implications for the low-temperature behavior, e.g. the decay 
of correlation functions, of the physical systems modeled by the Hamiltonian.
As stated before, Koma and Nachtergaele calculated the spectral gap of the
spin-1/2 chain in \cite{KN1}.  For spins greater than 1/2 Koma, Nachtergaele, 
and Starr
showed that there is a non-vanishing gap in one dimension in \cite{KNS}. These
results were extended by Caputo and Martinelli.  In \cite{CapMart2} they showed 
that the gap grows linearly with the spin.  This
confirmed a conjecture of Starr in \cite{St2}.  The techniques of Caputo and
Martinelli are limited
however, in the sense that they do not give an explicit value for a lower bound
of the spectral gap.  In Chapter \ref{chapter: xxz} we seek to remedy this in a
small way.  We present a method by Nachtergaele for estimating the gap of finite
chains in terms of the gap of the nearest-neighbor interaction 
$\frac{1}{\Delta}S^1_xS^1_y+\frac{1}{\Delta}S^2_xS^2_y+S^3_xS^3_y$.  The method
relies on an assumption (\ref{conjecture: epsilon}) that we believe to be 
satisfied for the spin-1 (and larger $J$) chain.
We have much numerical evidence to suggest that the assumption is satisfied, but
not an analytic proof of the fact.  For this reason the lower bound for the gap
of the spin-1 chain is only a conjecture, but we are confident that it is
accurate and will soon be rigorously verified.

As stated before the second result in this thesis concerns the  isotropic XY model 
in the prescence of an extermal magnetic field for dimensions 2 or greater.  The
Hamiltonian is given by
\be
H^{XY}_\Lambda =
-\sum_{\stackrel{\{x,y\}\in\Lambda}{|x-y|=1}}\frac{1}{2}\left(\s^1_x\s^1_y+\s^2_x\s^2_y\right) + h\sum_{x\in\Lambda}\frac{1}{2}\left(\idty-\s^3_x\right)
\ee
where the $\s^\alpha,\ \alpha=1,2,3$ are the Pauli spin matrices and $\Lambda$
is a finite subset of $\Ir^d$ with  $d\geq 2$.  In  a paper dated 1961,
\cite{LSM}, Lieb,
Shultz, and Mattis studied the general anisotropic one dimensional, spin-1/2, 
XY model without an external magnetic field ($h=0$.)  They rigorously established the 
equivalence of the XY model with a free fermion model via the Jordan-Wigner 
transformation, and proved the existence of long-range order for the anisotropic model
using a method they developed for calculating two point correlation functions.
In the isotropic case the order vanishes. They were also able
to establish the uniquess of a gapless ground state in the thermodynamic limit 
for the isotropic case. Later, Dyson, Lieb, and Simon, using reflection 
positivity, proved that the spin-1/2 XY model has a phase 
transition in three and higher dimensions \cite{DLS}. Finally, in 1988, Kennedy, Lieb, 
and Shastry demonstrated the existence of long-range order for the ferromagnetic 
XY model in dimensions two and greater, regardless of spin.  This then proved 
the existence of Bose-Einstein condensation for the hardcore gas on the same 
lattice at half filling \cite{KLS2}.

In Chapter \ref{chapter: xy} we show that for magnetic fields with $h>2d$, there
exists a range of non-zero temperatures where the XY model does not have a phase 
transition in two and higher dimensions.  This case is not treated in the above
papers since the prescence of the magnetic field destroys the reflection
positivity of the model.  In \cite{KT} Kennedy and Tasaki study a much larger
class of Hamiltonians, using similar techniques and with similar results.  At
this point, though, it remains unclear if there results cover the case with
which we are interested.
We employ a contour expansion, inspired by Kennedy in
\cite{Ken}, to show that the free energy of the system is analytic for some
values of inverse temperature $\beta$ and magnetic field $h>2d$.  Specifically
we are able to show that the weights of our contours satisfy a condition of
Koteck\'y and Preiss \cite{KP} which allows us to write the free energy using a cluster
expansion.

I have been supported in part by the National Science Foundation under
Grant No. DMS0070774, and VIGRE Grant No. DMS-0135345.

\newpage
\pagestyle{myheadings}
\chapter{XXZ Model}
\label{chapter: xxz}
\thispagestyle{myheadings} 
\markright{  \rm \normalsize CHAPTER 2. \hspace{0.5cm}
  XXZ Model }

For many spin systems interesting behaviors (e.g. phase transitions) only occur in
dimensions greater than one. However, in the case of the XXZ model, phase 
transitions do exist in dimension one.  This remarkable fact is closely
connected to the symmetry of the spin-1/2 chain under the action of the quantum
group $SU_q(2)$.

In \cite{KN1}, Koma and Nachtergaele calculated the exact spectral gap for the
spin-1/2 XXZ chain. The goal of this chapter is a lower bound for the spectral gap of the
1-dimensional spin-1 model.  This is obtained using a method developed by
Nachtergaele in \cite{Na2}, which was later improved by Spitzer and Starr in
\cite{SpSt}, for estimating the spectral gap of frustration free systems.

We will start by introducing the Hamiltonian for the XXZ chain. Since the method
we use to find a lower bound requires one to know the ground states of the finite volume
Hamiltonian, we will introduce these, and prove that they are indeed ground states 
for all values of spin.   For completeness, we also present a brief account of 
the infinite volume ground states, although this is a much more sophisticated subject.
Next we will give a proof, due to Koma and Nachtergaele, that the spectral gap
of the infinite volume spin-1/2 chain is given by $1-\Delta^{-1}$, where
$\Delta$ is the anisotropy parameter in the XXZ Hamiltonian.  In the last
section we introduce the method used to obtain the lower bound and present our
results.

%%%%%%%%%%%%%%%%%%%%%%%%%%%%%%%%%%%%%%%%%%%%%%%%%%%%%%%%%%%%%%%%%%%%%%%%%%%%%%%%
%%%%%%%%%%%%%%%%%%% SECTION: THE MODEL %%%%%%%%%%%%%%%%%%%%%%%%%%%%%%%%%%%%%%%%%
%%%%%%%%%%%%%%%%%%%%%%%%%%%%%%%%%%%%%%%%%%%%%%%%%%%%%%%%%%%%%%%%%%%%%%%%%%%%%%%%

\Section{The Model}

Initially we consider only finite intervals $\Lambda\subset\Ir$.  For each
$x\in\Lambda$ we let $\HH_x=\Cx^{(2J+1)}$ be the configuration space for the
individual spin at site $x$.  The Hilbert space for the entire chain is
given by $\HH_\Lambda = \bigotimes_{x\in\Lambda}\HH_x$.  The spin arises by way
of a (2J+1)-dimensional irreducible representation of $SU(2)$ on each $\HH_x$.  This 
is generated by three operators $S^1_x,\ S^2_x,\ S^3_x,$ which are determined relative
to a basis $\{\ket{-J}_x,\ \ket{-J+1}_x,\ldots,\ \ket{J}_x\}$ by
\be
\label{s3def}
S^3_x\ket{m}_x=m\ket{m}_x
\ee
and
\be
S^\pm_x\ket{m}_x=\sqrt{J(J+1)-m(m\pm{1})}\ket{m\pm{1}}_x
\ee
where $S_x^\pm = S_x^1\pm{i}{S}^2_x$.
The 1-dimensional spin-$J$ XXZ Hamiltonian is  then given by
\be
\label{XXZ1}
H^{XXZ}_L =
-\sum_{x=1}^{L-1}\frac{1}{\Delta}\left(S^1_xS^1_{x+1}+S^2_xS^2_{x+1}\right) + S^3_xS^3_{x+1}
\ee
where $\Delta>1$ is the anisotropy parameter.  The overall negative sign makes
this a ferromagnetic model, while the inclusion of the anisotropy forces low
energy states to have large regions in which spins are aligned along the 3-axis.  
Notice that in the limit $\Delta\rightarrow{1}$ we obtain the isotropic 
Heisenberg, or XXX, model, while as $\Delta\rightarrow\infty$ the model tends to
  an Ising like Hamiltonian.

As with many things in mathematics, the truly interesting behavior comes when we
add boundary terms.  To this end, we consider the so called ``kink'' boundary
conditions.  Let  $A(\Delta)=\sqrt{1-\Delta^{-2}}$ and define $H^{+-}_L$ by
\be
\label{XXZ Hamiltonian}
H^{+-}_L=H^{XXZ}_L + JA(\Delta)(S^3_L-S^3_1) + J^2(L-1)
\ee
The last term, $J^2(L-1)$, is added so that the ground state energy is zero.  If
we let 
\be
h_{x,x+1} = -\frac{1}{\Delta}\left(S^1_xS^1_{x+1}+S^2_xS^2_{x+1}\right) + S^3_xS^3_{x+1} + JA(\Delta)(S^3_{x+1} + S^3_x) + J^2
\ee
then $H^{+-}_L$ can be expressed as the sum over translates of a single two-body interaction:
\be
\label{hsum}
H^{+-}_L = \sum_{x=1}^{L-1}h_{x,x+1}
\ee 

The choice of $A(\Delta)$ is such that the spin-1/2 Hamiltonian will commute with
representations of the quantum group $SU_q(2)$ where $2\Delta = q+q^{-1}$ and
$q\in(0,1)$.  If we define the operator $K$ to be
\be
K = q^{-2S^3}
\ee
then $K,\ S^+,\ S^-$ satisfy the commutation relations
\be
[S^+,S^-] = \frac{K-K^{-1}}{q^{-1}-q}\ \ \mathrm{and}\ \ KS^{\pm}=q^{\mp{2}}S^{\pm}K
\ee
$K$ and $S^\pm$ are the generators of an irreducible representation of
$SU_q(2)$.
Amazingly, there is a way to define tensor products of representations of
$SU_q(2)$.  For a chain of $L$ sites we write

\bea
S^3_{[1,L]} &=&
\sum_{x=1}^L \idty_1\otimes\cdots\otimes\idty_{x-1}\otimes{S}^3_x\otimes\idty_{x+1}\otimes\cdots\otimes\idty_L\\
S^+_{[1,L]} &=& \sum_{x=1}^L K_1\otimes\cdots\otimes{K}_{x-1}\otimes{S}^+_x\otimes\idty_{x+1}\otimes\cdots\otimes\idty_L\\
S^-_{[1,L]} &=&
\sum_{x=1}^L
\idty_1\otimes\cdots\otimes\idty_{x-1}\otimes{S^-_x}\otimes{K^{-1}}_{x+1}\otimes\cdots\otimes{K^{-1}_L}
\eea

For spin-1/2 these operators commute with the XXZ Hamiltonian and were at the heart of the
initial discovery of the ground states of the XXZ model.  We will not use the
quantum group symmetry much here and only mention the existence of the symmetry
for completeness.  Someone who is interested in learning more about quantum
groups and there representations should consult the reference \cite{Kas}.

%%%%%%%%%%%%%%%%%%%%%%%%%%%%%%%%%%%%%%%%%%%%%%%%%%%%%%%%%%%%%%%%%%%%%%%%%%%%%%%%
%%%%%%%%%%% SECTION 1.1: FINITE VOLUME GS %%%%%%%%%%%%%%%%%%%%%%%%%%%%%%%%%%%%%%
%%%%%%%%%%%%%%%%%%%%%%%%%%%%%%%%%%%%%%%%%%%%%%%%%%%%%%%%%%%%%%%%%%%%%%%%%%%%%%%%

\Section{Finite Volume Ground States}
Perhaps the first, or most basic question, one can ask when given a Hamiltonian
like (\ref{XXZ Hamiltonian}) is: does it have any non-trivial ground states and
if so, what are they?  This question was first answered by Alcaraz, Salinas and
Wreszinski in \cite{ASW}.  As earlier indicated, they were able to realize the ground state of
(\ref{XXZ Hamiltonian}) in the spin-1/2 case as the highest weight irreducible
representation of the quantum group $SU_q(2)$.  They were able to generalize the
result for spins $J>1/2$.  Here we present their result, but the proof provided
is due to Starr in \cite{St2}.

%%%%%% THEOREM: ALCARAZ, SALINAS, WRESZINSKI %%%%%%%%%%%%%%%%%%%%%%%%%%%%%%%%%%%

\begin{theorem}(Alcaraz, Salinas, and Wreszinski)
For each eigenvalue $m$ of $\sum_{x=1}^LS^3_x$, $m\in\{-JL,\dots,+JL\}$ there is 
a unique (up to normalisation) state $\Psi_0^m$ in the sector of magnetization 
m, such that $H^{+-}_L\Psi_0^m=0$. $\Psi^m_0$ is given by
\be
\label{thmgs}
\Psi^m_0=\sum_{\stackrel{\{m_x\}\in\{-J,\dots, {J}\}^L}{\sum_xm_x=m}}{}\prod_{x=1}^{L}q^{-x(J-m_x)}{2J \choose J+m_x}^{1/2}\ket{\{m_x\}}
\ee
where $\ket{\{m_x\}}$ is just the simple tensor 
$$\ket{\{m_x\}} = \ket{m_1}\otimes\ket{m_2}\otimes\cdots\otimes\ket{m_L}$$ 
with respect to the natural basis for the spin operators given in (\ref{s3def}).
\end{theorem}
\begin{proof}
The goal is to calculate all states $\varphi$ such that $H^{+-}_L\varphi=0$, that
is we want to calculate $\mathrm{ker}\left(H^{+-}_L\right)$.  But by
(\ref{hsum}), $H^{+-}_L$ is the sum of non-negative operators, thus if
$\varphi\in\mathrm{ker}\left(H^{+-}_L\right)$ then $\varphi\in\mathrm{ker}\left(h_{x,x+1}\right)$
for $1\leq{x}\leq{L-1}$. This gives
\be
\mathrm{ker}\left(H^{+-}_L\right) = \bigcap_{x=1}^{L-1}\mathrm{ker}\left(h_{x,x+1}\right)
\ee
Models with this property are often referred to as {\bf{frustration free}}, since
any state that minimizes the energy of the global interaction also minizes the
energy of every local interaction and hence are not ``frustrated''.

We begin by calculating $\bigcap_{x=1}^{L-1}\mathrm{ker}\left(h_{x,x+1}\right)$
 for $J=1/2$.  For other values of $J$ we embed the spin-J chain into a longer 
spin-1/2 chain and use representation theory to obtain the desired result.
For $J=1/2$, $\HH_L$ is the $2^L$ dimensional space $\otimes_{x=1}^L\Cx^2_x$.
The simplest basis for $\HH_L$ is given by
$$\ket{\{m_x\}}=\bigotimes_{x=1}^L\ket{m_x}_x$$
where $\{m_x\}\in\{\pm{1/2}\}^L$.  Often we will use the convention that
$\ket{1/2}=\ket{\uparrow}$ and $\ket{-1/2}=\ket{\downarrow}$.  It is a short calculation to show that 
$$h_{x,x+1} = -\frac{1}{\Delta}\left(S^1_xS^1_{x+1}+S^2_xS^2_{x+1}\right) + S^3_xS^3_{x+1} + A(\Delta)(S^3_{x+1} - S^3_x) + J^2
$$
is the orthogonal projection onto the the vector 
\be
\ket{\xi}=\frac{1}{\sqrt{1+q^2}}\left(q\ket{\uparrow\downarrow}_{x,x+1}-\ket{\downarrow\uparrow}_{x,x+1}\right)
\ee
where $q$ is the unique solution to the equation $2\Delta=q+q^{-1}$ on the
interval $(0,1)$.  Thus, we can write
\be
\label{hproj}
h_{x,x+1} = \ket{\xi}\bra{\xi}
\ee

To motivate our argument for general $q$ consider the case when $q=1$.  Then $\ket{\xi}_{x,x+1}$ is the spin singlet, or the anti-symmetric
tensor, in $\HH_x\otimes\HH_{x+1}$.  If $\tau(x,x+1)\in\mathcal{S}_L$ is just
the transposition of $x$ and $x+1$ then again for $q=1$ we can write $h_{x,x+1}$ 
as
\be
h_{x,x+1} = \frac{1}{2}\left(\idty-P_{\tau(x,x+1)}\right)
\ee
where for $\pi\in\mathcal{S}_L$, $P_\pi$ represents the standard action of
$\mathcal{S}_L$ on $\HH_L$ given by
\be
\label{symrep}
P_\pi\bigotimes_{x=1}^{L}\ket{\phi_x}_x=\bigotimes_{x=1}^{L}\ket{\phi_{\pi^{-1}(x)}}_x
\ee
Hence, for the isotropic model ($\Delta=1)$, if $\psi$ is a ground state of 
$h_{x,x+1}$, then
$P_{\tau(x,x+1)}\psi$ is also a ground state for all $1\leq{x}\leq(L-1)$.  Since
the nearest-neighbor transpositions generate the entire group $\mathcal{S}_L$, we see
that kernel of the isotropic Hamiltonian is invariant under
permutations, i.e. it is made up of symmetric tensors.  This is quite evident
when it is noted that the XXX Hamiltonian has full $SU(2)$ symmetry, so that its
ground state coincides with the heighest weight irreducible representation of
$SU(2)$ in $\otimes_{x=1}^L\Cx^2$.

For $q\neq{1}$ we again realize the 
ground states as symmetric tensors, however not with respect to the standard 
action $P_\pi$. 
Let\\ $\mathbf{\omega}:\{\pm{1/2}\}^L\rightarrow\Ir$ be given by
$$\mathbf{\omega}(\{m_x\})=\sum_{x=1}^L-x\left(\frac{1}{2}-m_x\right)$$
and consider the basis for $\HH_L$ given by
$$\ket{\{m_x\}}^{(q)}=q^{\omega(\{m_x\})}\ket{\{m_x\}}$$
Now, let $P^{(q)}:\mathcal{S}_L\rightarrow{GL(\HH_L)}$ be the standard
representation of the symmetric group, but with respect to the basis
$\ket{m_x}^{(q)}$.  Thus we have 
$$P^{(q)}_\pi\ket{\{m_x\}}^{(q)} = \ket{\{m_{\pi^{-1}(x)}\}}^{(q)}$$
but with respect to the standard basis we get
\beann
P^{(q)}_\pi\ket{\{m_x\}} &=& P^{(q)}_\pi\left(q^{-\omega(\{m_x\})}\ket{\{m_x\}}^{(q)}\right)\\
&=& q^{-\omega(\{m_x\})}\ket{\{m_{\pi^{-1}(x)}\}}^{(q)}\\
&=& q^{\omega(\{m_{\pi^{-1}(x)}\})-\omega(\{m_x\})}\ket{\{m_{\pi^{-1}(x)}\}}
\eeann
This formula looks a little complicated, but it is exactly what we want.  If we
restrict our attention to the space $\HH_x\otimes\HH_{x+1}$ then the symmetric
group only consists of two elements, the identity $e$ and the transpostition
$\tau(x,x+1)\equiv\tau$.  A short calculation shows that 
\beann
P^{(q)}_\tau\ket{\uparrow\uparrow}=\ket{\uparrow\uparrow} && P^{(q)}_\tau\ket{\uparrow\downarrow}=q\ket{\downarrow\uparrow}\\
P^{(q)}_\tau\ket{\downarrow\downarrow}=\ket{\downarrow\downarrow} && P^{(q)}_\tau\ket{\downarrow\uparrow}=q^{-1}\ket{\uparrow\downarrow}
\eeann
Then we can immediately write down the symmetric tensors as
$$
\ket{\uparrow\uparrow},\ \ket{\downarrow\downarrow},
\frac{\ket{\uparrow\downarrow} + q\ket{\downarrow\uparrow}}{\sqrt{1+q^2}}
$$
while the anti-symmetric tensor is given by
$$\frac{\ket{\uparrow\downarrow} - q\ket{\downarrow\uparrow}}{\sqrt{1+q^2}}$$
Thus, by (\ref{hproj}), $h_{x,x+1}$ is the orthogonal projection onto the
anti-symmetric tensor of the representation $P^{(q)}$, and hence we write
$$h_{x,x+1} = \frac{1}{2}\left(\idty - P^{(q)}_{\tau(x,x+1)}\right)$$
where the anisotropy parameter $\Delta$ in $h_{x,x+1}$ and $q\in(0,1)$ are 
related by $2\Delta = q + q^{-1}$.  So, just as in the isotropic case, we are
able to conclude that any frustration free state of $H^{+-}_L$ is invariant
under nearest-neighbor transpositions.  But nearest-neighbor transpostions
generate the symmetric group, so the ground states of $H^{+-}_L$ are
symmetric with respect to the representation $P^{(q)}$.  Since $P^{(q)}$ is
identical to
the standard representation with respect to the basis $\ket{\{m_x\}}^{(q)}$ we
can immediately write down the symmetric states with respect to this basis:
$$\Psi^{m}_0 = \sum_{\stackrel{\{m_x\}\in\{\pm{1/2}\}^L}{\sum{m_x}=m}}\ket{\{m_x\}}^{(q)}$$
where $m\in\{-L/2, (-L/2+1),\ldots, L/2\}$.  Expressing this in terms of the 
standard basis gives
\bea
\Psi^{m}_0 &=& \sum_{\stackrel{\{m_x\}\in\{\pm{1/2}\}^L}{\sum{m_x}=m}}q^{\omega(\{m_x\})}\ket{\{m_x\}} \nonumber\\
&=& \sum_{\stackrel{\{m_x\}\in\{\pm{1/2}\}^L}{\sum{m_x}=m}}q^{\sum-x(1/2-m_x)}\ket{\{m_x\}} \nonumber\\
&=&
\sum_{\stackrel{\{m_x\}\in\{\pm{1/2}\}^L}{\sum{m_x}=m}}\prod_{x=1}^Lq^{-x(1/2-m_x)}\ket{\{m_x\}} \label{hlfgs}
\eea
which is the correct formula for $J=1/2$.

For $J>1/2$ we let $\HH^{(J)}_x=\Cx^{2J+1}$ as above.  $\HH^{(J)}_x$ can be
realized as the highest-weight space in the representation
$\otimes_{j=1}^{2J}\Cx^2_{x_j}$.  In other words $\HH^{(J)}_x$ consists of the
symmetric tensors in $\otimes_{j=1}^{2J}\HH^{(1/2)}_{x_j}$.  But we have $L$
copies of $\HH^{(J)}_x$ thus the space we want is
\be
\mathbb{H}^{(J)}_L = \bigotimes_{i=1}^L\left(\bigotimes_{j=1}^{2J}\HH^{(1/2)}_{x_j,i}\right)_i
\ee

Again, suppose that $P:\mathcal{S}_{2J}\rightarrow{GL}(\HH^{(1/2)}_{2J})$ is the standard
representation of $\mathcal{S}_{2J}$ on $\HH^{(1/2)}_{2J}$  with respect to the basis $\ket{\{m_{x_j}\}}$. 
Let $\mathcal{S}_{2J}^{\times{L}}$ be the direct product of $L$ copies of the
symmetric group on $2J$ elements.   Define a representation $\Pi$ of
$\mathcal{S}_{2J}^{\times{L}}$ onto $\mathbb{H}^{(J)}_L$ by
$$\Pi_{(\pi_1,\ldots,\pi_L)}\bigotimes_{i=1}^{L}\ket{\{m_{x_j,i}\}}_i =
\bigotimes_{i=1}^LP_{\pi_i}\ket{\{m_{x_j,i}\}}_i$$
so that the $i^{\mathrm{th}}$ copy of $\mathcal{S}_{2J}$
acts in the standard way on the $i^{\mathrm{th}}$ copy of
$\otimes_{j=1}^{2J}\HH^{(1/2)}_{x_j}$ in $\mathbb{H}^{(J)}_L$.
Let $\mathrm{Sym}_\Pi[\mathbb{H}^{(J)}_L]\subset \mathbb{H}^{(J)}$ denote the
subspace of $\mathbb{H}^{(J)}$ which is invariant under the action of $\Pi$.
This space coincides with $\HH^{(J)}_L$ because
\bea
\mathrm{Sym}_\Pi[\mathbb{H}^{(J)}_L] &=&
\mathrm{Sym}_\Pi\left[\bigotimes_{i=1}^L\left(\bigotimes_{j=1}^{2J}\HH^{(1/2)}_{x_j,i}\right)_i\right]\\
&=&
\bigotimes_{i=1}^L\left(\mathrm{Sym}_{P}\left[\bigotimes_{j=1}^{2J}\HH^{(1/2)}_{x_j,i}\right]\right)_i
\eea
since $\Pi$ only acts on individual factors in $\mathbb{H}^{(J)}_L$ via $P$.  But
$$\mathrm{Sym}_P\left[\bigotimes_{j=1}^{2J}\HH^{(1/2)}_{x_j,i}\right] = \HH^{(J)}_i$$ 
and hence
$$
\mathrm{Sym}_\Pi[\mathbb{H}^{(J)}_L] = \bigotimes_{i=1}^L\HH^{(J)}_i = \HH^{(J)}_L
$$
Next, we define a spin-1/2 Hamiltonian on $\mathbb{H}^{(J)}$ which reduces to the
Hamiltonian $H^{+-,(J)}_L$.  We consider $L$ distinct and disjoint copies of the
set $\{1,2,\ldots,2J\}$ which we denote by $[1,2J]_i$, $i=1,\ldots,L$.  Then
define $H^{+-,(1/2)}_{(2J_1,\ldots,2J_L)}$ by
\be
H^{+-,(1/2)}_{(2J_1,\ldots,2J_L)} = \sum_{i=1}^{L-1}\sum_{\stackrel{(x_i,x_{i+1})}
{x_i\in[1,2J]_i,\ x_{i+1}\in[1,2J]_{i+1}}} h_{x_i,x_{i+1}}
\ee
$H^{+-,(1/2)}_{(2J_1,\ldots,2J_L)}$ is clearly invariant under the action of
$\Pi$.
Therefore, setting $\OP_{{Sym}_\Pi}$ as the orthogonal projection onto
$\mathrm{Sym}_\Pi[\mathbb{H}^{(J)}_L]$, gives 
$$H^{+-,(J)}_L=\OP_{{Sym}_\Pi}H^{+-,(1/2)}_{(2J_1,\ldots,2J_L)}\OP_{{Sym}_\Pi}$$.
However, the construction of $\Pi$ was such that $H^{+-,(1/2)}_{(2J_1,\ldots,2J_L)}$ commutes
with the entire action, thus
$$H^{+-,(J)}_L = H^{+-,(1/2)}_{(2J_1,\ldots,2J_L)}\OP_{{Sym}_\Pi}$$
This means that $\mathrm{ker}(H^{+-,(J)}_L) =
\mathrm{ker}(H^{+1,(1/2)}_{2JL})\cap\mathrm{ran}(\OP_{Sym_\Pi})$
However, by the argument for the spin-1/2 case, the ground state of
$H^{+-,(1/2)}_{(2J_1,\ldots,2J_L)}$ is invariant under the action $P^{(q)}$.  Therefore 
$$\mathrm{ker}(H^{+-,(1/2)}_{(2J_1,\ldots,2J_L)})\subset\mathrm{ran}(\OP_{Sym_\Pi})$$
which implies
$$\mathrm{ker}(H^{+-,(J)}_L) =
\mathrm{ker}(H^{+-,(1/2)}_{(2J_1,\ldots,2J_L)})\cap\mathrm{ran}(\OP_{Sym_\Pi}) = \mathrm{ker}(H^{+-,(1/2)}_{(2J_1,\ldots,2J_L)})$$
that is, the ground state of $H^{+-,(J)}_L$ is just the ground state of
$H^{+-,(1/2)}_{(2J_1,\ldots,2J_L)}$. But the formula for the ground states of
$H^{+-,(1/2)}_{(2J_1,\ldots,2J_L)}$ is given by (\ref{hlfgs}).  Writing
these in terms of symmetric tensors; doing so yields
\beann
\Psi^m_0 &=&
\sum_{\stackrel{\{m_{x_j}\}\in\{\pm{1/2}\}^{2JL}}{\sum_jm_{x_j}=m}}\prod_{j=1}^{2JL}q^{-j(1/2-m_{x_j})}\ket{\{m_{x_j}\}}\\
&=& \sum_{\stackrel{\{m_{x_j}\}\in\{\pm{1/2}\}^{2JL}}{\sum_jm_{x_j}=m}}\bigotimes_{j=1}^{2JL}q^{-j(1/2-m_{x_j})}\ket{m_{x_j}}_j\\
\eeann
Next, we rewrite the single sum as two sums; first over the L chains of length
$2J$, and then over the bases of the individual chains.  This gives
\beann
\Psi^m_0
&=&
\sum_{\stackrel{\{m_x\}\in\{-J,\ldots{J}\}^L}{\sum_xm_x=m}}\bigotimes_{x=1}^Lq^{-x(J-m_x)}\left(\sum_{\stackrel{\{b_j\}_x\in\{\pm{1/2}\}}{\sum_jb_{j,x}=m_x}}\bigotimes_{j=1}^{2J}\ket{b_{j,x}}_j\right)\\
&=&
\sum_{\stackrel{\{m_x\}\in\{-J,\ldots{J}\}^L}{\sum_xm_x=m}}\bigotimes_{x=1}^Lq^{-x(J-m_x)}{2J \choose J+m_x}^{1/2}\ket{m_x}_x\\
&=& \sum_{\stackrel{\{m_x\}\in\{-J,\ldots{J}\}^L}{\sum_xm_x=m}}\prod_{x=1}^Lq^{-x(J-m_x)}{2J \choose J+m_x}^{1/2}\ket{\{m_x\}}
\eeann
which is the same as (\ref{thmgs}).
\end{proof}

%%%%%%%%%%%%%%%%%%%%%%%%%%%%%%%%%%%%%%%%%%%%%%%%%%%%%%%%%%%%%%%%%%%%%%%%%%%%%%%%
%%%%%%%%%%%%%% SECTION 1.2: INFINITE VOLUME GROUND STATES %%%%%%%%%%%%%%%%%%%%%%
%%%%%%%%%%%%%%%%%%%%%%%%%%%%%%%%%%%%%%%%%%%%%%%%%%%%%%%%%%%%%%%%%%%%%%%%%%%%%%%%

\section{Infinite Volume Ground States}

In infinite volume the problem of determining the ground states for the
Hamiltonian is much more subtle; even defining the Hamiltonian turns out not to
be as simple as one might think.  For $\Lambda\subset\Ir$, $|\Lambda|<\infty$ the
algebra of observables  is $\obs_\Lambda\equiv\mathcal{B}(\HH_\Lambda)$.  In
infinite volume the algebra of quasi-local observables $\obs_\infty$ is defined by
$$\obs_\infty \equiv
\overline{\left(\bigcup_{\stackrel{\Lambda\subset\Ir}{|\Lambda|<\infty}}\obs_\Lambda\right)}$$
where the completion is taken in the operator norm topology.
Note that if $\Lambda'\subset\Lambda$ then there is a
natural way to embed $\obs_{\Lambda'}$ as a subalgebra of $\obs_\Lambda$, namely
\be
\label{infgs}
\obs_{\Lambda'}\ni{A'} \hookrightarrow A'\otimes\idty_{\Lambda\setminus\Lambda'}\in \obs_\Lambda
\ee
In the infinite volume $H_\Ir$ is determined by a derivation $\delta$ on 
$\obs_{\infty}$.  For any $A\in\obs_\Lambda$, $|\Lambda|<\infty$
$$\delta(A) = \lim_{\Lambda'\nearrow\Ir}\left[H_{\Lambda'},A\right]$$
A state, $\omega$, on $\obs_\infty$ is a positive (normalized) linear
functional.  A ground state is one that satisfies the local stability property
$$\omega(A^*\delta(A))\geq{0}\ \ \ \ A\in\obs_\Lambda;\ |\Lambda|<\infty$$
This condition becomes much more intuitive if we go to the GNS representation
corresponding to the state $\omega$: $(\pi, \HH, \Omega)$.  For the XXZ 
Hamiltonian with $+-$ boundary condition the GNS Hilbert space, $\HH_{+-}$, can be understood
in a simple way.  Let $\Omega^{+-}$ be given by
\be
\label{Omegadef}
\Omega^{+-}=\bigotimes_{x\in\Ir}\Omega^{+-}_x,\ \ \ \ \
\Omega^{+-}_x=\left\{\begin{array}{c}
\ket{\uparrow}\ \ x\leq{0} \\
\ket{\downarrow} \ \ x\geq{1}
\end{array}
\right.
\ee
Then the set of vectors of the form
\be
\label{+-GNS core}
\psi = A\Omega^{+-}\ \ \mathrm{where}\  A\in\bigcup_{\Lambda\subset\Ir}\obs_\Lambda,\ |\Lambda|<\infty 
\ee
is a dense subset of $\HH_{+-}$. For a local observable $A$, the stability 
condition (\ref{infgs}) becomes
\beann
\omega(A^*\delta(A)) &=& \ip{\Omega^{+-}}{\pi(A^*)\delta(\pi(A))\Omega^{+-}}\\
&=& \ip{\Omega^{+-}}{\pi(A^*)}{[\pi(H),\pi(A)]\Omega^{+-}}\\
&=& \ip{\Omega^{+-}}{\pi(A^*)\pi(H)\pi(A)\Omega^{+-}} - \ip{\Omega^{+-}}{\pi(A^*)\pi(A)\pi(H)\Omega^{+-}}\\
&=& \ip{\Omega^{+-}}{\pi(A^*)\pi(H)\pi(A)\Omega^{+-}}\\ 
&=& \bra{\pi(A)\Omega^{+-}}\ \pi(H)\ \ket{\pi(A)\Omega^{+-}}\ \geq\ 0
\eeann
Thus we see that (\ref{infgs}) designates a
ground state as one for which any local change coincides with some increase in
energy; which is exactly what a ground state should be.

A closely related, but stronger, notion in the infinite volume is that of a zero
energy state.  For a model like the XXZ, we can define this in the following
way.  Let $H_\Lambda = \sum_{\{x,x+1\}\subset\Lambda}h_{x,x+1}$ where
$h_{x,x+1}$ is the translate of some nearest-neighbor interaction $h_{1,2}$
for all $(x,x+1)$. Then $\omega$ is a zero energy state iff
$$\omega(h_{x,x+1}) = \min_{\omega'\in\obs^*_\infty}\omega'(h_{x,x+1})$$
for all $\{x,x+1\}\subset\Ir$ where here the states are all normalized.  This is
also called frustration free as in the finite volume case.  If
$A\in\obs_{[1,L]}$, then
\beann
\omega(A^*\delta(A)) &=&
\lim_{\Lambda\nearrow\Ir}\omega(A^*(H_\Lambda{A}-AH_\Lambda))\\
&=& \sum_{x=0}^{L}\omega(A^*(h_{x,x+1}A-Ah_{x,x+1}))\\
&=& \sum_{x=0}^{L}\omega(A^*(h_{x,x+1}-\lambda_0)A)\\
&=& \sum_{x=0}^{L}\omega_A(h_{x,x+1}-\lambda_0) \geq 0
\eeann
where $\lambda_0$ is the smallest eigenvalue of $h_{x,x+1}$  and
$\omega_A(\cdot) = \omega(A^*\cdot{A})$.  Therefore, if $\omega$ is a zero energy
state then it is also a ground state, however the reverse is not in general
true.  For the XXZ chain Gottstein and Werner found all zero energy states in
\cite{GW}.  Their results and techniques were quite general and will not be
presented here,
although they obtained very nice results connecting zero energy states and VBS
states.  Moreover, they predicted correctly that for the XXZ and
XXX chains the zero energy states are the only ground states.  Recalling the
definition of $\Omega^{+-}$  in (\ref{Omegadef}), the zero energy states for the
XXZ Hamiltonian are given by
\be
\label{infinite volume zero energy state}
\Psi_0(n)=\sum_{k=0}^\infty\sum_{\stackrel{\{x_1,\ldots,x_k\}\subset\Ir_{\leq{0}}}{\{y_1,\ldots,y_{n+k}\}\subset\Ir_{\geq{1}}}} q^{-(x_1+\cdots+x_k)+(y_1+\cdots+y_{n+k})}\prod_{j=1}^kS^-_{x_j}\prod_{j=1}^{n+k}S^+_{y_j}\Omega^{+-}
\ee
It was shown by Matsui in \cite{Mat1} that the zero energy states defined above along with
those obtained from $-+$, $++$ and $--$ boundary conditions account for all
ground states of the XXZ chain.  Later, Koma and Nachtergaele in \cite{KN3} gave an
alternative  proof of this fact with the added result that their techniques could be
applied to the XXX chain as well. This is a major difference between Koma and 
Nachtergaele's and Matsui's results. Their results not only apply to spin-$1/2$, 
but to all $J$. 

%%%%%%%%%%%%%%%%%%%%%%%%%%%%%%%%%%%%%%%%%%%%%%%%%%%%%%%%%%%%%%%%%%%%%%%%%%%%%%%%
%%%%%%%%%%%%%%% SECTION 1.3: THE SPECTRAL GAP %%%%%%%%%%%%%%%%%%%%%%%%%%%%%%%%%
%%%%%%%%%%%%%%%%%%%%%%%%%%%%%%%%%%%%%%%%%%%%%%%%%%%%%%%%%%%%%%%%%%%%%%%%%%%%%%%%

\Section{The Spectral Gap}

%%%%%%%%%%%%%%%% SUBSECTION 1.3.1: GAP OF SPIN-1/2 CHAIN %%%%%%%%%%%%%%%%%%%%%%%
\subsection{Spin-1/2}
The spectral gap of the spin-$1/2$ XXZ chain was first calculated by Koma and
Nachtergaele in \cite{KN1}.  Their proof is a truly beautiful piece of work that
makes use of standard tools of statistical mechanics, such as the transfer matrix,
while also employing newer phenomena such as the quantum group symmetry of the 
XXZ model.  Together with S. Starr they later showed for all $J$ there is a 
non-zero spectral gap in the thermodynamic limit, although they did not provide 
an estimate \cite{KNS}.  For completeness, we provide the proof of the existence of the gap in
the spin-$1/2$ case.

%%%%%%%%%%%%%%%% THEOREM: KOMA NACTHERGAELE  GAP OF SPIN 1/2 CHAIN %%%%%%%%%%%%%

For $\alpha,\beta\in\{+,-\}$, let $\Omega^{\alpha\beta}$ be the generalizations
of $\Omega^{+-}$ given in (\ref{Omegadef}), that is 
\be
\Omega^{\alpha\beta} = \bigotimes_{x\in\Ir}\Omega^{\alpha\beta}(x)\ \ \mathrm{where}\ \  
\Omega^{\alpha\beta}(x) = \left\{\begin{array}{ccc}
\ket{\alpha} && \mathrm{if}\ x\leq{0}\\
\ket{\beta} && \mathrm{if}\ x\geq{1}
\end{array}\right.
\ee
and similiarly the GNS space $\HH_{\alpha\beta}$ has a dense set of vectors of
the form
\be
\label{GNS core}
\psi = A\Omega^{\alpha\beta}\ \ \ \ \mathrm{where}\ \ 
A\in\bigcup_{\Lambda\subset\Ir}\obs_\Lambda,\ \ \ |\Lambda|<\infty
\ee
The cyclic vector of the GNS representation, $\Omega^{\alpha\beta}_{GNS}$ is
then given by a generalization of equation (\ref{infinite volume zero energy
state})

\be
\label{cyclic GNS states}
\Omega^{\alpha\beta}_{GNS} =
Z^{-1}(q)\sum_{k=0}^\infty\sum_{\stackrel{\{x_1,\ldots,x_k\}\subset\Ir_{\leq{0}}}{\{y_1,\ldots,y_k\}\subset\Ir_{\geq{1}}}}q^{\sum_{j=1}^k(y_j-x_j)}\prod_{j=1}^kS^{\beta}_{x_j}S^{\alpha}_{y_j}\Omega^{\alpha\beta}
\ee
where $Z^{-1}(q)$ is a normalization factor.  The Hamiltonian 
$H^{\alpha\beta}_{GNS}$ is then defined by the equation
\be
H^{\alpha\beta}_{GNS}A\Omega^{\alpha\beta}_{GNS} =
\lim_{\Lambda\nearrow\Ir}[H^{\alpha\beta}_\Lambda,A]\Omega^{\alpha\beta}_{GNS}
\ee
Note that the set of vectors in (\ref{GNS core}) are in the domain of
$H^{\alpha\beta}_{GNS}$.   

\begin{theorem}[Koma and Nachtergaele] 
\label{KNgap}
For all $\Delta>1$ and for any choice of GNS sector ($(+-),\ (-+),\ (++),\ (--)$)
let $\Gamma^{\alpha\beta}$ be the set of all $\lambda\in\Rl$ such that
\be
\ip{\Omega^{\alpha\beta}}{A^*\left(H^{\alpha\beta}_{GNS}\right)^3A\Omega^{\alpha\beta}} \geq \lambda \ip{\Omega^{\alpha\beta}}{A^*\left(H^{\alpha\beta}_{GNS}\right)^2A\Omega^{\alpha\beta}}
\ee
for all $A\in\cup_{\Lambda}\obs_\Lambda$.  Then
\be
\sup \Gamma^{\alpha\beta} = 1-\Delta^{-1}
\ee
or, in other words, if $\gamma_{\alpha\beta}$ is the spectral gap of
$H^{\alpha\beta}_{GNS}$ then
\be
\gamma_{\alpha\beta} = 1 - \Delta^{-1}.
\ee
\end{theorem}
The proof of theorem \ref{KNgap} consists of two main parts: first an estimate
on a lower bound for $\gamma$ and then an argument to show that this is also an upper bound.
\\

\begin{proof}
In order to bound $\gamma$ from below by $1-\Delta^{-1}$ we show for all local
observables $A$
\be
\ip{\Omega^{\alpha\beta}}{A^*\left(H^{\alpha\beta}_{GNS}\right)^3A\Omega^{\alpha\beta}} \geq (1-\Delta^{-1}) \ip{\Omega^{\alpha\beta}}{A^*\left(H^{\alpha\beta}_{GNS}\right)^2A\Omega^{\alpha\beta}}
\ee
that is, for every $\psi$ in the range of $H^{\alpha\beta}_{GNS}$ 
\be
\ip{\psi}{H^{\alpha\beta}_{GNS}\psi} \geq (1-\Delta^{-1})\ip{\psi}{\psi}
\ee
But, if $A\in\obs_\Lambda$ with $\Lambda$ finite, then
\be
\ip{\Omega^{\alpha\beta}}{A^*\left(H^{\alpha\beta}_{GNS}\right)^kA\Omega^{\alpha\beta}} = \ip{\Omega^{\alpha\beta}}{A^*\left(H^{+-}_{\Lambda+\{-1,0,1\}}\right)^kA\Omega^{\alpha\beta}}
\ee
 The operator
$A^*\left(H^{+-}_{\Lambda+\{-1,0,1\}}\right)^kA$ is in 
$\obs_{\Lambda+\{-1,0,1\}}$, and thus we can write
\be
\ip{\Omega^{\alpha\beta}}{A^*\left(H^{+-}_{\Lambda+\{-1,0,1\}}\right)^kA\Omega^{\alpha\beta}} = \mathrm{Tr}\left[\rho_{\Lambda+\{-1,0,1\}}A^*\left(H^{+-}_{\Lambda+\{-1,0,1\}}\right)^kA\right]
\ee
 where $\rho_{\Lambda+\{-1,0,1\}}$ is a density matrix.  Hence, we have shown 
 the infinite volume gap $\gamma$ is greater than or equal to $1-\Delta^{-1}$ if, 
for all finite volumes $\Lambda$, the spectral gap of $H^{\alpha\beta}_\Lambda$ is 
 greater than $1-\Delta^{-1}$.  This is the result of the following proposition. 
 
%%%%%% PROPOSITION: GAP = 1-\DELTA^(-1)cos(PI/L) %%%%%%%

\begin{proposition} 
\label{KNfingap}
Let $\gamma_L$ denote the spectral gap of $H^{+-}_L$.  Then for all $\Delta\geq{1}$
and $L\geq{2}$ we have
\be
\gamma_L=1-\Delta^{-1}\cos(\pi/L)
\ee
\end{proposition} 
To prove Propositon \ref{KNfingap} we calculate the first non-zero eigenvalue in
the subspace of $\HH_L$ corresponding to eigenvectors of $S^3_{\mathrm{tot}}$
with eigenvalue $L/2-1$, i.e. the sector with one downspin, and show that this
is also a lower bound for the smallest non-zero eigenvalue in all other sectors. 

Let $D_x=S^-_x\ket{\uparrow}^L$ where $\ket{\uparrow}^L =
\otimes_{x=1}^L\ket{\uparrow}_x$.  Then if $\psi$ is a vector in the sector with 
one downspin 
\be
\label{der1}
\psi=\sum_{x=1}^{L}a_xD_x
\ee
Now suppose that $\psi$ is an eigenvector with eigenvalue $\mathcal{E}$, then
\be
\label{der2}
H^{+-}\psi = \sum_{x=1}^La_xH^{+-}_LD_x = \mathcal{E}\psi
\ee
On the other hand computing $H^{+-}_L$ gives
\bea
H^{+-}_LD_1 &=& \frac{1+A(\Delta)}{2}D_1-\frac{1}{2\Delta}D_2 \nonumber\\
H^{+-}_LD_x &=& D_x-\frac{1}{2\Delta}(D_{x-1}+D_{x+1})\ \ \ \mathrm{for}\ 2\leq{x}\leq{L} \nonumber\\
H^{+-}_LD_L &=& \frac{1-A(\Delta)}{2}D_L-\frac{1}{2\Delta}D_{L-1} \label{der3}
\eea
Putting (\ref{der1})-(\ref{der3}) together we get the equations
\bea
a_{x+1} &=& 2\Delta(1-\mathcal{E})a_x-a_{x-1} \ \ \mathrm{for}\
2\leq{x}\leq{L-1},\label{rcr1}\\
a_2 &=& 2\Delta(1/2 +A(\Delta)/2 - \mathcal{E})a_1\label{rcr2}\\
a_{L-1} &=& 2\Delta(1/2-A(\Delta)/2-\mathcal{E})a_L\label{rcr3}
\eea
Notice that if $a_1=0$ then the above implies $\psi=0$, so we can assume from 
now on that $a_1\neq{0}$.  Equation (\ref{rcr1}) can be written as
\be
\left(\begin{array}{c}
a_{x+1}\\
a_x
\end{array}\right) = T\left(\begin{array}{c}
a_{x}\\
a_{x-1}
\end{array}\right)
\ee
where $T=\left(\begin{array}{cc} 2\Delta(1-\mathcal{E})&-1\\ 1 &
0\\\end{array}\right)$ is called the transfer matrix.  Iterating this gives
\be
\label{trans1}
\left(\begin{array}{c}
a_{L}\\
a_{L-1}
\end{array}\right) = T^{L-2}\left(\begin{array}{c}
a_{2}\\
a_{1}
\end{array}\right)
\ee
So taking into account (\ref{rcr2}), (\ref{rcr3}), and (\ref{trans1}) we arrive
at 
\be
\label{trans2}
a_L\left(\begin{array}{c}
1\\
\Delta[1 - A(\Delta) - \mathcal{E}]\end{array}\right) =
a_1T^{L-2}\left(\begin{array}{c}
\Delta[1 + A(\Delta) - \mathcal{E}]\\
1
\end{array}\right)
\ee
Equation (\ref{trans2}) can be solved using the eigenvectors and eigenvalues of
$T$.  The characteristic equation of $T$ is 
\be
\label{Tchar}
\lambda^2-2\Delta(1-\mathcal{E})\lambda+1 = 0
\ee
which has roots
\be
\label{Teigs}
\lambda_{\pm} = \Delta(1-\mathcal{E})\pm\sqrt{\Delta^2(1-\mathcal{E})^2-1}
\ee
Initially, let us consider the case when $T$ does not have degenerate
eigenvalues, that is $\Delta(1-\mathcal{E})\neq{\pm{1}}$.  The eigenvectors of $T$ are
\be
\label{Tvects}
e_{\pm} = \left(\begin{array}{c}
\lambda_{\pm}\\
1
\end{array}\right)
\ee
We want to solve equation (\ref{trans2}) so we must decompose each vector into
a sum of $e_+$ and $e_-$.  Projecting the vector on the left  of (\ref{trans2})
onto $e_\pm$ gives
\be
\label{leftvec}
\left(\begin{array}{c}
1\\
\Delta[1 - A(\Delta) - \mathcal{E}]\end{array}\right) = \mu_+\lambda_-e_+ +
\mu_-\lambda_+e_-
\ee
where
\be
\label{mudef}
\mu_\pm=\frac{1}{2}\left[1\pm\Delta\frac{A(\Delta)+\mathcal{E}}{\sqrt{\Delta^2(1-\mathcal{E})^2-1}}\right]
\ee
and the one on the right hand side of (\ref{trans2}) gives
\be
\label{rightvec}
\left(\begin{array}{c}
\Delta[1 + A(\Delta) - \mathcal{E}]\\
1
\end{array}\right) = \eta_+e_+ + \eta_-e_-
\ee
where
\be
\label{etadef}
\eta_\pm = \frac{1}{2}\left[1\pm\Delta\frac{A(\Delta)-\mathcal{E}}{\sqrt{\Delta^2(1-\mathcal{E})^2-1}}\right]
\ee
Now we can substitute (\ref{leftvec}) and (\ref{rightvec}) back into
(\ref{trans2}) to get
\bea
a_L(\mu_+\lambda_-e_+ +\mu_-\lambda_+e_-) &=& a_1T^{L-2}(\eta_+e_+ +\eta_-e_-)\\
&=& a_1(\eta_+T^{L-2}e_+ + \eta_-T^{L-2}e_-\\
&=& a_1(\eta_+\lambda_+^{L-2} + \eta_-\lambda_-^{L-2}e_-)
\eea
But $e_+$ and $e_-$ are linearly independent thus we get the equations
\bea
\frac{a_L}{a_1}\mu_+ &=& \frac{\lambda_+^{L-2}}{\lambda_-} = \lambda_+^{L-1}\label{rat1}\\
\frac{a_L}{a_1}\mu_- &=& \frac{\lambda_-^{L-2}}{\lambda_+} =
\lambda_-^{L-1}\label{rat2}
\eea
where we have used the fact that $\lambda_+\lambda_-=1$.  Simplifying
(\ref{rat1}) and (\ref{rat2}) further gives 
\be
\label{lambdarat}
\lambda^{2L-2} = \frac{\mu_+}{\mu_-}\times\frac{\eta_-}{\eta_+}
\ee
where we have assumed that $\eta_-\neq{0}$.  This assumption is not restrictive
since 
\bea
\eta_-=0 &\iff&
\frac{1}{2}\left[1-\Delta\frac{A(\Delta)-\mathcal{E}}{\sqrt{\Delta^2(1-\mathcal{E})^2-1}}\right] = 0\\
&\iff& \Delta\left[A(\Delta)-\mathcal{E}\right] =
\sqrt{\Delta^2(1-\mathcal{E})^2-1}\\
&\iff& \mathcal{E}\sqrt{\Delta^2-1} = \Delta\mathcal{E}\\
&\iff& \mathcal{E} = 0
\eea
thus $\eta_-=0$ corresponds to the ground state in the $L/2-1$ sector.
Moreover, $\eta_-\neq{0}$ guarantees that all $\mu_\pm$ and $\eta_\pm$ are
non-zero, so we see that equation (\ref{lambdarat}) is indeed valid in the
setting we want.  Using (\ref{mudef}), (\ref{etadef}), and (\ref{Teigs}), we can
write
\be
\label{murat}
\frac{\mu_+}{\mu_-} =
\frac{\Delta+\sqrt{\Delta^2-1}}{\lambda_+}\times\frac{\lambda_+-(\Delta-\sqrt{\Delta^2-1})}{\lambda_+-(\Delta+\sqrt{\Delta^2-1})}
\ee
and 
\be
\label{etarat}
\frac{\eta_-}{\eta_+} = \frac{\Delta-\sqrt{\Delta^2-1}}{\lambda_+}\times\frac{\lambda_+-(\Delta+\sqrt{\Delta^2-1})}{\lambda_+-(\Delta-\sqrt{\Delta^2-1})}
\ee
Together with (\ref{lambdarat}) this gives 
\be
\lambda_+^{2L}=1
\ee
which implies 
\be
\lambda_+=e^{i\pi\ell/L}\ \ \ \ell\in\Ir
\ee
Returning now to (\ref{Teigs}) and solving for $\mathcal{E}_L(\ell)$ gives
\be
\label{excitedenergies}
\mathcal{E}_L(\ell) = 1 - \frac{\lambda_++\lambda_-}{2\Delta} = 1 -
\Delta^{-1}\cos(\pi\ell/L) \ \ \ell = 1, 2, \ldots,L-1
\ee
where the condition on $\ell$ comes from the assumption that (\ref{Tchar}) has
non-degenerate roots. Furthermore, we see that equation (\ref{excitedenergies})
gives $L-1$ distinct eigenvalues, which gives us a total of $L$ distinct
eigenvalues of $H^{+-}_L$ in the $(L/2-1)$ sector when we add the ground state
energy $\mathcal{E}=0$.  More to the point, there are not any solutions with
$\Delta(1-\mathcal{E})=\pm{1}$.  

The above proves Proposition \ref{KNfingap}, that is in the $(L/2-1)$ sector the
smallest non-zero eigenvalue is given by $1-\Delta^{-1}\cos(\pi/L)$.  To finish
the proof of Theorem \ref{KNgap} we must show that this is same for all other
sectors.  In order to accomplish this we utilize the quantum group symmetry of
the model.   Note that any irreducible representation of $SU_q(2)$ of dimension
$L-1$ must intersect the sector with one overturned spin.  Thus if the first
excited state, $\varphi^m_1$, lies in an $L-1$ irreducible representation for any choice of
magnetization $m,\ 1<{m}<L$, we can raise it to a state, $\phi^m_1$,  in  the $L/2-1$ sector by 
way of the quantum group raising operator.  The state $\phi^m_1$ has energy greater than or equal
to $1-\Delta^{-1}\cos(\pi/L)$ by the argument above.  However, since the quantum
group commutes with the Hamiltonian $H^{+-}_L$, $\phi^m_1$ has the same
energy as the original ground state, $\varphi^m_1$,  and Theorem \ref{KNgap} follows.  Thus we
must show that the lowest excitations lie in the $L-1$ dimensional irreducible
representations of $SU_q(2)$.  To this end we present the following lemma without
proof.  The proof makes use of straighforward applications of the quantum group
symmetry as well techniques similar to those in Section \ref{subsect:spin1}.

%%%%%%%%%%%%% LEMMA: SMALLEST EXCITATIONS LIE IN L-1 IRREPS %%%%%%%%%%%%%%%%%%%%
Consider an arbitrary spin chain of length $n$. Let $\HH_{S^3_{tot}\geq{nJ-s}}$ 
be the span of eigenvectors of $S^3_{tot}$ with eigenvalue greater than $nJ-s$ 
(less than $nJ$.) Let $H_n=\sum_{x=1}^{n-1}h_{x,x+1}$ be an $SU_q(2)$ invariant Hamiltonian on a spin $J$
chain with $n$ sites.    Let $\gamma_n$ denote the spectral gap of $H_n$
and let $\epsilon_n^{(s)}$ be given by
\be
\epsilon_n^{(s)} =
\min_{\stackrel{0\neq{\psi}\perp\operatorname{ker}H_n}{\psi\in\HH_{S^3_{tot}\geq{nJ-s}}}}\frac{\ip{\psi}{H_n\psi}}{\|\psi\|^2}
\ee
Then the following holds
\begin{lemma}[Koma and Nachtergaele \cite{KN1}]
\label{L-1 irreps have lowest excitation}
Let $H_L$ be as above. Suppose that $h_{x,x+1}\geq{0}$ and that
$\mathrm{ker}(H_L)$ is non-empty.  Furthermore, suppose that $\mathrm{ker}(H_n)$
coincides with the irreducible representation of $SU_q(2)$ with maximal spin
(=$nJ$,)for $2\leq{n}\leq{L}$.  If
\be
\epsilon_n^{(2J)}\leq\epsilon_{n+1}^{(2J)}
\ee 
for all $2\leq{n}\leq{L-1}$, then
\be
\gamma_L = \epsilon^{(2J)}_L
\ee.
\end{lemma}
The calculation with the transfer matrix, combined with Lemma \ref{L-1 irreps
have lowest excitation}, completes the proof of Proposition \ref{KNfingap}.
Thus, for all $L\geq{2}$, the spectral gap of $H^{+-}_L$ is given by
\be
\label{bound on gamma}
\gamma_L = 1 - \Delta^{-1}\cos(\pi/L) \geq 1 - \Delta^{-1} 
\ee
Thus $1-\Delta^{-1}$ is indeed a lower bound for $\gamma$.  

In order to complete 
the proof of Theorem \ref{KNgap}, we must show that $\gamma$ is less than or
equal to $1-\Delta^{-1}$.  This is a little subtle;  upon initial inspection one
can see that the gaps for $H^{++}_{GNS}$ and $H^{--}_{GNS}$  are the same by
spin flip symmetry, and that the gaps for $H^{+-}_{GNS}$ and $H^{-+}_{GNS}$ are the
same by left-right symmetry.  However, the ground state of $H^{++}_{GNS}$ is 
non-degenerate but the ground state of $H^{+-}_{GNS}$ is infinitely degenerate, so it 
is not so clear that the gap for $H^{++}_{GNS}$ and the gap for $H^{+-}_{GNS}$ are the same.  
It is known that the translation invariant states are weak limits of the (anti-)
kink states. Thus, 
\begin{multline}
\inf_{\Lambda,A\in\obs_\Lambda}
\frac{\ip{\Omega^{\alpha\beta}}{A^*(H^{+-}_{\Lambda+\{-1,0,1\}})^3A\Omega^{\alpha\beta}}}{\ip{\Omega^{\alpha\beta}}{A^*(H^{+-}_{\Lambda+\{-1,0,1\}})^2A\Omega^{\alpha\beta}}}  \\
\leq\  \inf_{\Lambda,A\in\obs_\Lambda} \lim_{n\rightarrow\pm\infty}
\frac{\ip{\Omega^{\alpha\beta}}{\tau_n(A^*(H^{+-}_{\Lambda+\{-1,0,1\}})^3A)\Omega^{\alpha\beta}}}{\ip{\Omega^{\alpha\beta}}{\tau_n(A^*(H^{+-}_{\Lambda+\{-1,0,1\}})^2A)\Omega^{\alpha\beta}}}\\
=\ \inf_{\Lambda,A\in\obs_\Lambda}
\frac{\ip{\Omega^{++}}{A^*(H^{+-}_{\Lambda+\{-1,0,1\}})^3A\Omega^{++}}}{\ip{\Omega^{++}}{A^*(H^{+-}_{\Lambda+\{-1,0,1\}})^2A\Omega^{++}}}
\end{multline}
so $\gamma^{+-}\leq\gamma^{++}$ and hence we must only show that $1-\Delta^{-1}$
is greater that or equal to $\gamma^{++}$ in order to finish the proof of 
Theorem \ref{KNgap}.  Towards this end we start with the variational priciple
\be
\gamma^{++} =
\inf_{\stackrel{0\neq\psi\perp\mathrm{ker}H^{++}_{GNS}}{\psi\in\mathrm{dom}H^{++}_{GNS}}}\frac{\ip{\psi}{H^{++}_{GNS}\psi}}{\|\psi\|^2}
\ee
In addition we note that since $h_{x,x+1}\geq{0}$, for any sequence of
sets $\Lambda_1\subset\Lambda_2\subset\cdots, n\in\Nl$,
$\HH_{++}\supset\mathrm{ker}H^{++}_{\Lambda_1}\supset\mathrm{ker}H^{++}_{\Lambda_2}\supset\cdots$.
So, if $\psi$ must be in a space orthogonal to $\mathrm{ker}H^{++}_{GNS}$, $\psi$
need only satisfy $\psi\perp\mathrm{ker}H^{++}_\Lambda$ for some particularly
well chosen $\Lambda$.

Let $\Lambda=[1,n]\subset\Ir$.  Define the spin wave operators, $X_k,\ \ k\in\frac{2\pi}{n}\{1,2,\ldots,n-1\}$ by
\be
X_k = \frac{1}{\sqrt{n}}\sum_{x=1}^ke^{ikx}S^-_x
\ee
where we multiply by $1/\sqrt{n}$ so that
\be
\ip{\Omega^{++}}{X_k^*X_l\Omega^{++}} = \delta_{k,l}
\ee
To get the upper bound we take $\psi = (c_1X_{k_1}+c_2X_{k_2})\Omega^{++}$.
Notice that $\psi$ is in the subspace of $\HH_{++}$ with one overturned spin;
specifically, $\sum_{x=1}^nS^3_x\psi = (n/2-1)\psi$.  But for each eigenvalue $m$ of
$\sum_{x=1}^nS^3_x$ there is exactly one ground state $\psi^m_0$ of
$H^{+-}_{[1,n]}$  such that $\sum_{x=1}^nS^3_x\psi^m_0=m\psi^m_0$.  Therefore
for fixed distinct $k_1$ and $k_2$ we can pick $c_1,\ c_2$ such that
$\psi\perp\psi^{(n-2)/2}_0$.  But this implies that 
\be
\label{gap upper bound}
\inf_{n,k_1,k_2}\sup_{c_1,c_2} \frac{\ip{\psi}{H^{++}_{GNS}\psi}}{\|\psi\|^2} \geq
\gamma^{++}
\ee
However, we will show that
\be
\label{bound is right}
\inf_{n,k_1,k_2}\sup_{c_1,c_2} \frac{\ip{\psi}{H^{++}_{GNS}\psi}}{\|\psi\|^2} =
1-\Delta^{-1}
\ee
which together with (\ref{gap upper bound}) establishes $1-\Delta^{-1}$ as an
upper bound for $\gamma^{++}$.  

For $1\leq{x},y\leq{n}$ define $T_{x,y}$ by
\bea
T_{x,y} &=& \ip{\Omega^{++}}{S^+_xH^{+-}_{[0,n+1]}S^-_y\Omega^{++}}\nonumber\\
&=& \frac{1}{2\Delta}(2\Delta\delta_{x,y}-\delta{x,y-1}-\delta_{x,y+1})
\eea
Then a simple, messy computation shows that the $\sup_{c_1,c_2}$ in (\ref{bound
is right}) gives the norm of the matrix $M(n,k_1,k_2)$  whose matrix elements\\
$M(n,k_1,k_2)_{i,j}=M_n(k_i,k_j)$ are given by
\be
M_n(k,l) = \frac{1}{n}\sum_{x,y=1}^ne^{-ikx}T_{x,y}e^{ily} =
\delta_{k,l}(1-\Delta^{-1}\cos{k})+\frac{e^{i(l-k)}}{2\Delta{n}}
\ee
for $k,l=2\pi{m}/n$.  Thus we have 
\be
\inf_{n,k_1,k_2}\sup_{c_1,c_2} \frac{\ip{\psi}{H^{++}_{GNS}\psi}}{\|\psi\|^2} =
\inf_{n,k_1,k_2} \|M(n,k_1,k_2\| = 1-\Delta^{-1}
\ee
which finishes the proof that $\gamma^{++}\leq{1}-\Delta^{-1}$, which in turn
gives us Theorem \ref{KNgap}

\end{proof}

%%%%%%%%%%%%%%%%% SUBSECTION: SPIN-1 GAP %%%%%%%%%%%%%%%%%%%%%%%%%%%%%%%%%%%%%%%
\subsection{Spin-1}
\label{subsect:spin1}
For spins greater than one-half, the quantum group symmetry, which was
instrumental in the calculation of the spectral gap in the spin-1/2 model, is
lost.  Although this makes questions about the gap for spins greater than 
one-half harder, it does not
make them untreatable.  In \cite{KNS}  Koma, Nachtergaele, and Starr showed that
for all values of spin, there is a non-vanishing gap in the thermodynamic limit.
 They also conjectured that the gap grew linearly with the spin.  This
 prediction was confirmed in \cite{CapMart2} by Caputo and Martinelli as a corollary to results on
 interacting particle systems.  Using the equivalence of the Markov generators
 for certain reaction-diffusion processes and Hamiltonians of quantum spin
 models,
 %\cite{Alcaraz}
 Caputo and Martinelli were able to show that for all values of anisotropy
 parameter $\Delta>1$ there is a constant $\delta$ such that 
 $$\delta{J}\leq\mathrm{gap}\left(H^{+-,(J)}_L\right)\leq\delta^{-1}J$$
 for all values of spin-$J$ and chains of length $L\geq{2}$.  Here we use
 techniques developed by Nachtergaele to give a concrete lower bound for the
 gap in the case $J=1$.  We begin by introducing the techniques for a slightly
 more general class of Hamiltonians than the XXZ; specifically the method will
 work for any Hamiltonian with frustration free ground states, so we only assume
 this.  The method presented here has been significantly improved by Spitzer and Starr in \cite{SpSt}, but there method is a bit more complicated, so for simplicity we present the original method here.  We then give the estimate and then seek to justify it through a series
 of Lemmas.

%%%%%%%%%%%%%%%%% SUBSUBSECTION: MARTINGALE METHOD %%%%%%%%%%%%%%%%%%%%%%%%%%%%%
\subsubsection{The Martingale Method}

We start with a spin chain on $L$ sites and label the Hilbert space as 
$\HH_L=\left(\Cx^d\right)^{\otimes{L}}$.  Let $h_{1,2}\geq{0}$ be a two-site 
Hamiltonian acting non-trivially on the first two sites, and let $h_{x,x+1}$ 
denote the translate of $h_{1,2}$ acting non-trivially on the factors at sites 
$\{x,x+1\}$.  We consider the Hamiltonian 
$$H_L=\sum_{x=1}^{L-1}h_{x,x+1}$$
and assume that the kernel of $H_L$ is non-trivial. 
If $1\leq{a}<b\leq{L}$ then we denote by $G_{[a,b]}$ the orthogonal projection 
onto the kernel of $\sum_{x=a}^{b-1}h_{x,x+1}$, and use the convention that 
$G_{\{x\}}=\idty$.  More generally, for $\Lambda\subset[1,L]$ we set $G_\Lambda$
to be the orthogonal projection onto
$$\mathrm{ker}\sum_{x,\{x,x+1\}\subset\Lambda}h_{x,x+1}$$
Then the $G_\Lambda$ satisfy
\bea
G_{\Lambda_1}G_{\Lambda_2} &=& G_{\Lambda_2}G_{\Lambda_1}=G_{\Lambda_2}\ \ \mathrm{if}\ \Lambda_1\subset\Lambda_2 \label{Gprop1}\\
G_{\Lambda_1}G_{\Lambda_2} &=& G_{\Lambda_2}G_{\Lambda_1}\ \ \mathrm{if}\ \ \Lambda_1\cap\Lambda_2=\emptyset \label{Gprop2}
\eea
The above follows directly from the assumption that $h_{1,2}\geq{0}$, and thus we have for any $\Lambda\subset[1,L]$
$$\mathrm{ker}\ H_\Lambda\ =\ \bigcap_{x,\{x,x+1\}\subset\Lambda}\mathrm{ker}\ h_{x,x+1}$$
Moreover if $\gamma$ is the smallest non-zero eigenvalue of $h_{1,2}=H_2$ then 
\be {\label{Gprop3}}
h_{x,x+1}\geq\gamma(\idty-G_{[x,x+1]})
\ee
Next define the family of orthogonal projections $\{E_n:1\leq{n}\leq{L}\}$ by
\be{\label{Edef}}
E_n = \left\{\begin{array}{ll}
\idty-G_{[1,2]}& \mathrm{if}\ n=1\\
G_{[1,n-1]}-G_{[1,n]}& \mathrm{if}\ 2\leq{n}\leq{L-1}\\
G_{[1,n]}& \mathrm{if}\ n=L
\end{array}\right.
\ee
Using (\ref{Gprop1}) and (\ref{Gprop2}) it is clear that the $E_n's$ are self-adjoint and satisfy
\bea
E_nE_m &=& \delta_{n,m}E_n \label{Eprop1}\\
\sum_{n=1}^LE_n &=& \idty \label{Eprop2}
\eea

\begin{theorem}[The Martingale Method]
Assume that $H$ is a Hamiltonian satisfying the properties above.  Furthermore, assume that for each $n,\ 1\leq{n}\leq{L-1}$, 
\be
\|G_{[n,n+1]}E_{n}\|<1/\sqrt{2} \label{assumption}
\ee
Then for all $\psi$ such that $G_{[1,n]}\psi={0}$ 
\be
\ip{\psi}{H\psi}\geq\gamma(1-\sqrt{2}{\varepsilon})^2\|\psi\| \label{gapesti}
\ee  
where $\varepsilon=\max_n\|G_{[n,n+1]}E_{n}\|$ and $\gamma$ is the spectral gap 
of the nearest-neighbor Hamiltonian $h_{1,2}$.
\end{theorem}

\begin{proof}
Take $\psi$ such that  $G_{[1,n]}\psi={0}$.  By (\ref{Eprop1}) and (\ref{Eprop2})
$$\|\psi\|^2=\sum_{n=1}^L\|E_n\psi\|^2$$
but $E_n=G_{[1,n]}$ so
$$\|\psi\|^2=\sum_{n=1}^{L-1}\|E_n\psi\|^2$$

We then use $\ip{\psi}{h_{x,x+1}\psi}$ to estimate $\|E_n\psi\|$ by writing:
\benn
\|E_n\psi\|^2 = \ip{\psi}{(1-G_{[n,n+1]})E_n\psi}+\ip{\psi}{G_{[n,n+1]}E_n\psi}  
\eenn
then, we insert the resolution of the identity given by (\ref{Eprop2}) into the second term noting that $G_{[n,n+1]}\psi=0$ to get
\be
\|E_n\psi\|^2 = \ip{\psi}{(1-G_{[n,n+1]})E_n\psi}+\ip{\psi}{\sum_{m=1}^{L-1}E_mG_{[n,n+1]}E_n\psi}
\ee
The second term simplifies further.
\beann
E_mG_{[n,n+1]} &=& (G_{[1,m]}-G_{[1,m+1]})G_{[n,n+1]}\\
&=& G_{[n,n+1]}(G_{[1,m]}-G_{[1,m+1]})\\
&=& G_{[n,n+1]}E_m
\eeann
for $m<(n-1)$ or $m>n+1$ by (\ref{Gprop2}).  But if $m\neq{n}$, then $E_mE_n=0$ by (\ref{Eprop1}) so that
\bea
\lefteqn{\|E_n\psi\|^2 = \ip{\psi}{(\idty-G_{[n,n+1]})E_n\psi}} \hspace{3cm}\nonumber \\
&&{}+\ip{\psi}{(E_{n-1}+E_n)G_{[n,n+1]}E_n\psi} \\
&=& \ip{\psi}{(\idty-G_{[n,n+1]})E_n\psi}\nonumber\\
&&{}+\ip{(E_{n-1}+E_n)\psi}{G_{[n,n+1]}E_n\psi} \label{ensi}
\eea
We now estimate both terms in (\ref{ensi}) using the identity
$$|\ip{\varphi_1}{\varphi_2}|\leq\frac{1}{2c}\|\varphi_1\|^2+\frac{c}{2}\|\varphi_2\|^2$$
for $c>0$.  Hence
\bea
\|E_n\psi\|^2 &=& \ip{(\idty-G_{[n,n+1]})\psi}{E_n\psi}+\ip{(E_{n-1}+E_n)\psi}{G_{[n,n+1]}E_n\psi}\nonumber\\
&\leq& \frac{1}{2c_1}\ip{\psi}{(\idty-G_{[n,n+1]})\psi}+\frac{c_1}{2}\ip{\psi}{E_n\psi}\\
&& {}+\frac{1}{2c_2}\ip{\psi}{E_nG_{[n,n+1]}E_n\psi}+\frac{c_2}{2}\ip{\psi}{(E_{n-1}+E_n)^2\psi} \nonumber\\
&=& \frac{1}{2c_1}\ip{\psi}{(\idty-G_{[n,n+1]})\psi}+\frac{c_1}{2}\|E_n\psi\|^2 \label{gdesti}\\
&& {}+\frac{1}{2c_2}\|G_{[n,n+1]}E_n\psi\|^2+\frac{c_2}{2}\|(E_{n-1}+E_n)\psi\|^2 \nonumber
\eea
The first term in (\ref{gdesti}), $\frac{1}{2c_1}\ip{\psi}{(\idty-G_{[n,n+1]})\psi}$,  can be estimated using (\ref{Gprop3}) and gives
\be
\frac{1}{2c_1}\ip{\psi}{(\idty-G_{[n,n+1]})\psi} \leq \frac{1}{2c_1\gamma}\ip{\psi}{h_{n,n+1}\psi} \label{1term}
\ee
The third term is then treated using (\ref{assumption}) to get
\be
\frac{1}{2c_2}\|G_{[n,n+1]}E_n\psi\|^2 \leq \frac{\varepsilon^2}{2c_2}\|E_n\psi\|^2 \label{3term}
\ee
Finally the last term can be split up as 
\be
\frac{c_2}{2}\|(E_{n-1}+E_n)\psi\|^2 = \frac{c_2}{2}(\|E_{n-1}\psi\|^2+\|E_n\psi\|^2) \label{4term}
\ee
since $E_{n-1}$ and $E_n$ are mutually orthogonal projections.
Thus by (\ref{gdesti})-(\ref{4term}) we have
\beann
\|E_n\psi\|^2 &\leq& \frac{1}{2c_1\gamma}\ip{\psi}{h_{n,n+1}\psi} + \frac{c_1}{2}\|E_n\psi\|^2 + \frac{\varepsilon^2}{2c_2}\|E_n\psi\|^2 \nonumber \\
&& {} + \frac{c_2}{2}(\|E_{n-1}\psi\|^2+\|E_n\psi\|^2)
\eeann
Summing over $n$ gives
\benn
\|\psi\|^2 \leq \frac{1}{2c_1\gamma}\ip{\psi}{H\psi} + \left(\frac{c_1}{2}+\frac{\varepsilon^2}{2c_2}+c_2\right)\|\psi\|^2
\eenn
where we note that there is not an $E_0$ term and that $E_L\psi=0$.  Rearranging this expression gives the inequality
\be
\ip{\psi}{H\psi} \geq \gamma{c}_1\left(2-c_1-\frac{\varepsilon^2}{c_2}-2c_2\right)\|\psi\|^2 \label{nslvineq}
\ee
Maximizing ${c}_1\left(2-c_1-\frac{\varepsilon^2}{c_2}-2c_2\right)$ over $c_1>0$, $c_2>0$ gives $c_1=1-\varepsilon\sqrt{2}$ and $c_2=\varepsilon/\sqrt{2}$.  Subsituting these into (\ref{nslvineq}) gives (\ref{gapesti}).

\end{proof}

%%%%%%%%%%%%%%%%%%%%%%%%%%%%%%%%%%%%%%%%%%%%%%%%%%%%%%%%%%%%%%%%%%%
%%%%%%%%%%%%%%%%%%%%%%%%%%%%%%%%%%%%%%%%%%%%%%%%%%%%%%%%%%%%%%%%%%%

\subsubsection{Main Result and Conjectures}

We present here an explicit lower bound of the spin-1 XXZ Hamiltonian
based on the Martingale Method given above.  The result relies on an, as yet,
unverified assumption (\ref{assumption}).  Since the projections $G_\Lambda$ and
$E_n$ all commute with $S^3_{\mathrm{tot}}=\sum_{x=1}^LS_x^3$, we can first 
project into each sector of $S^3_{\mathrm{tot}}$, and then check to see if
assumption (\ref{assumption}) is true.  This is the strategy we employ and it is
the substance of Lemma \ref{lemma: epsilon L} and conjecture (\ref{conjecture:
epsilon}).  We are able to verify that for a spin-1 chain of $L+1$ sites that
(\ref{assumption}) holds in the sector of magnetization $L$.  For the remainder
of the sectors we are unable to verify (\ref{assumption}) analytically, but we
do have numerical evidence to support it.

We proceed as follows: In Theorem \ref{theorem: lower bound} we apply the
Martingale Method to get a lower bound for
the gap of the spin-1 XXZ Hamiltonian as a function of the two-site gap $\gamma$.
In Lemma \ref{lemma: two-site gap} we calculate the gap, $\gamma$, of the two site
Hamiltonian.  In Lemma \ref{lemma: epsilon L} we calculate $\|G_{[L,L+1]}E_L\|$ 
in the sector with one overturned spin and show that it is less than 
$1/\sqrt{2}$ for all $L$, followed by the 
conjecture, along with numerical support, that the maximum $\|G_{[L,L+1]}E_L\|$ 
is achieved in the sector with one overturned spin.  

\begin{theorem}
\label{theorem: lower bound}
Consider the spin-1 XXZ chain on $L+1$ sites, and assume that Conjecture
\ref{conjecture: epsilon} holds.
Let $\psi\neq{0}$ be such that 
$\mathrm{G}_{[1,L+1]}\psi=0$.  Then
\be
\inf_{\psi}\frac{\ip{\psi}{H_{L+1}\psi}}{\|\psi\|^2}\geq\gamma\left(1-\sqrt{\frac{2q^2}{1+q^2}\cdot\frac{1-q^{2L}}{1-q^{2L+2}}}\right)^2
\ee 
where $\gamma$ is the gap of the two-site Hamiltonian given by 
$\frac{5}{2}-\sqrt{\frac{9}{4}-\frac{2}{\Delta^2}}$. 

\end{theorem}

\begin{lemma}[$\gamma$: gap of $h_{1,2}$]
\label{lemma: two-site gap}
Let $h_{1,2}$ be given by
$$h_{1,2} = -\frac{1}{\Delta}\left(S_1^1S_2^1+S_1^2S_2^2\right) - S_1^3S_2^3 - A(\Delta)\left(S_1^3 - S_2^3\right) + \idty$$ 
and let $\psi\neq{0}$.  Then
\be
{\label{2gap}}
\gamma=\min_{\psi:G_{[1,2]}\psi=0}\frac{\ip{\psi}{h_{1,2}\psi}}{\|\psi\|^2}=\frac{5}{2}-\sqrt{\frac{9}{4}-\frac{2}{\Delta^2}}
\ee

\end{lemma}

\begin{proof}
This is really more of a calculation than a proof.  The calculation proceeds in the following way:
Let $P_m$ be the projection onto the subspace of $\HH_2$ with total $S^3$ component equal to $m$.  
We calculate $\gamma_m=\mathrm{gap}\left(P_mh_{1,2}P_m\right)$ for $m=-2,-1,\ldots,2$,  
which is just  the first non-zero eigenvalue since $h_{1,2}$ has a non-trivial kernel. 
Then $\gamma$ is just the minimum of the $\gamma_m$.
For $m=\pm{2}$ the subspaces are one dimensional and both vectors are ground states.  
For $m=\pm{1}$ both $P_{\pm{1}}h_{1,2}P_{\pm{1}}$ can be expressed as the matrix
\be
P_{\pm{1}}h_{1,2}P_{\pm{1}} = \left(\begin{array}{cc}
1+A(\Delta) & -\frac{1}{\Delta}\\
-\frac{1}{\Delta} & 1-A(\Delta)
\end{array}\right)
\ee
where $A(\Delta)=\sqrt{1-1/\Delta^2}$.  The matrix has eigenvalues 0 and 2, so we see that 
\be
\label{gamma1}
\gamma_{\pm{1}}=2
\ee
For $m=0,{}P_{0}h_{1,2}P_{0}$ the matrix is 
\be
P_{0}h_{1,2}P_{0} = \left(\begin{array}{ccc}
2(1-A(\Delta)) & -\frac{1}{\Delta} & 0 \\
-\frac{1}{\Delta} & 1 & -\frac{1}{\Delta}\\
0 & -\frac{1}{\Delta} & 2(1+A(\Delta))
\end{array}\right)
\ee
Here the matrix also has a zero eigenvalue; when $\lambda$ is factored out of the characteristic equation we get 
$$\lambda^2-5\lambda+(\frac{2}{\Delta^2}+4) = 0$$
This has roots 
$$\lambda = \frac{5}{2} \pm \sqrt{\frac{9}{4}-\frac{2}{\Delta^2}}$$
so we get 
\be
\label{gamma0}
\gamma_0 = \frac{5}{2} - \sqrt{\frac{9}{4}-\frac{2}{\Delta^2}}
\ee
Taking the minimum of (\ref{gamma1}) and (\ref{gamma0}) we see that

\be
\gamma = \frac{5}{2} - \sqrt{\frac{9}{4}-\frac{2}{\Delta^2}}
\ee
\end{proof}

We now show that in the sector with magnetization $L$,
$\|G_{[L,L+1]}E_L\|<1/\sqrt{2}$ for all $q\in(0,1)$.  This is the first step in
verifying (\ref{assumption}) for the spin-1 chain.  We mention that the 
calculation given here for the spin-1 case immediately generalizes to higher
$J$, but we do not present that here.
Let $P_m$ be the orthogonal projection onto the sector with total $S^3$ component 
equal to $m$.  Then define 
$\varepsilon_m\equiv\|P_m\mathrm{G}_{[L,L+1]}\mathrm{E}_LP_m\|$. 

\begin{lemma}
\label{lemma: epsilon L}
$\varepsilon_L = 
\left[\frac{(1-q^{2L})}{(1-q^{2L+2})}\cdot\frac{q^2}{1+q^2}\right]^{1/2}<\left[\frac{q^2}{1+q^2}\right]^{1/2}$.  
Thus for all values of $q$ in $(0,1)$, $\varepsilon_L<1/\sqrt{2}$.
\end{lemma}

\begin{proof}
To begin with we write
\be\psi^L_{L+1}=\psi^{L-1}_L\otimes\ket{\uparrow} + \sqrt{2}q^{-L-1}\psi^L_L\otimes\ket{0}\ee
 where $\psi^M_N$ is a spin-1 groundstate on a chain of length $N$ with total
 $S^3$ component $M$. Define
$$
\varphi=\frac{\sqrt{2}q^{-L-1}}{\|\psi^{L-1}_L\|^2}\psi^{L-1}_L\otimes\ket{\uparrow}-\psi^L_L\otimes\ket{0}.
$$
Now $\varepsilon_L=\frac{\|\mathrm{G}_{[L,L+1]}\varphi\|}{\|\varphi\|}$, so we calculate $\mathrm{G}_{[L,L+1]}\varphi$
\beann
\mathrm{G}_{[L,L+1]}\varphi &=& \mathrm{G}_{[L,L+1]}\left(\frac{\sqrt{2}q^{-L-1}}{\|\psi^{L-1}_L\|^2}\psi^{L-1}_L\otimes\ket{\uparrow}-\psi^L_L\otimes\ket{0}\right)\\
&=& \frac{\sqrt{2}q^{-L-1}}{\|\psi^{L-1}_L\|^2}\psi^{L-2}_{L-1}\otimes\ket{m=2}+\frac{1}{\sqrt{1+q^2}}\left(\frac{2q^{-2L}}{\|\psi^{L-1}_L\|^2}-1\right)\psi^{L-1}_{L-1}\otimes\ket{m=1}\\
\\
&=& \frac{\sqrt{2}q^{-L-1}}{\|\psi^{L-1}_L\|^2}\psi^{L-2}_{L-1}\otimes\ket{m=2} - \frac{1}{\sqrt{1+q^2}}\frac{\|\psi^{L-2}_{L-1}\|^2}{\|\psi^{L-1}_L\|^2}\psi^{L-1}_{L-1}\otimes\ket{m=1}\\
\eeann
where $\ket{m=k}$ is the ground state of $H_{[L,L+1]}$ such that
$$(S^3_L+S^3_{L+1})\ket{m=k} = k\cdot\ket{m=k}$$
Calculating $\|\mathrm{G}_{[L,L+1]}\varphi\|^2$, we get
\beann
\|\mathrm{G}_{[L,L+1]}\varphi\|^2 &=& \frac{1}{\|\psi^{L-1}_L\|^4}\left[2q^{-2L-2}\|\psi^{L-2}_{L-1}\|^2+\frac{\|\psi^{L-2}_{L-1}\|^4}{1+q^2}\right]\\
\\
&=& \frac{\|\psi^{L-2}_{L-1}\|^2}{\|\psi^{L-1}_L\|^4}\frac{\left(2q^{-2L-2}+2q^{-2L}+\|\psi^{L-2}_{L-1}\|^2\right)}{1+q^2}
\eeann
But $2q^{-2L-2}+2q^{-2L}+\|\psi^{L-2}_{L-1}\|^2=\|\psi^{L}_{L+1}\|^2$ so we have
\be
\|\mathrm{G}_{[L,L+1]}\varphi\|^2 = \frac{\|\psi^{L-2}_{L-1}\|^2\|\psi^{L}_{L+1}\|^2}{\|\psi^{L-1}_L\|^4(1+q^2)}
\ee
On the other hand 
\benn
\|\varphi\|^2 = \frac{\|\psi^{L}_{L+1}\|^2}{\|\psi^{L-1}_L\|^2}
\eenn
so that
\be
\label{epsilonsquared}
\varepsilon_L^2=\frac{\|\mathrm{G}_{[L,L+1]}\varphi\|^2}{\|\varphi\|^2}=\frac{\|\psi^{L-2}_{L-1}\|^2}{\|\psi^{L-1}_L\|^2}\cdot\frac{1}{1+q^2}.
\ee
Substituting $\|\psi^{L-1}_L\|^2=2\sum_{x=1}^Lq^{-2x}$ into (\ref{epsilonsquared}) 
gives the expression
\be
\varepsilon_L^2=\frac{(1-q^{2L})}{(1-q^{2L+2})}\cdot\frac{q^2}{1+q^2}.
\ee
Hence $\varepsilon_L^2 < 1/2$ since for $q\in(0,1)$, $(1-q^{2L})/(1-q^{2L+2}) < 1$ and $q^2/(1+q^2) < 1/2$.
\end{proof}

What remains in order to verify assumption (\ref{assumption}), is to show that 
in sectors with $m<L$, $\varepsilon_m \leq \varepsilon_L$.  As stated
previously this is really the hard part.  At this point we are unable to prove
this analytically, but we state the conjecture here and provide a little
justification, both numeric and heuristic.

\begin{conjecture}
\label{conjecture: epsilon}
$\varepsilon_m\leq\varepsilon_L$ for all $m$ such that $-L\leq{m}<L$.
\end{conjecture}

Some of the strongest evidence we have for this conjecture is numeric.  We show
here the values of $\varepsilon_m$ in all sectors for a chain of length eight
and various values of $\Delta$.

\begin{figure}[h]
\begin{center}
\includegraphics[scale=0.3]{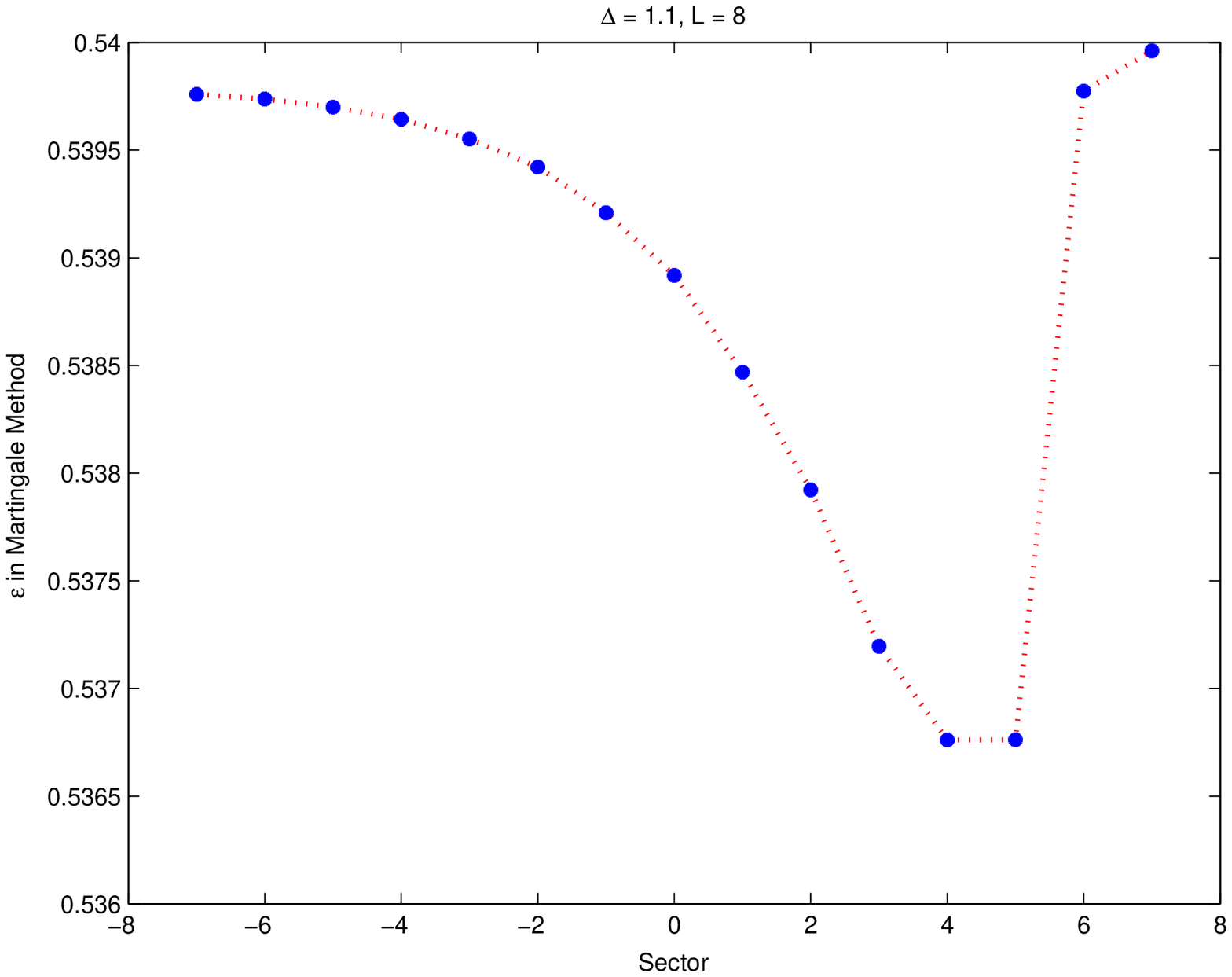}\
\includegraphics[scale=0.3]{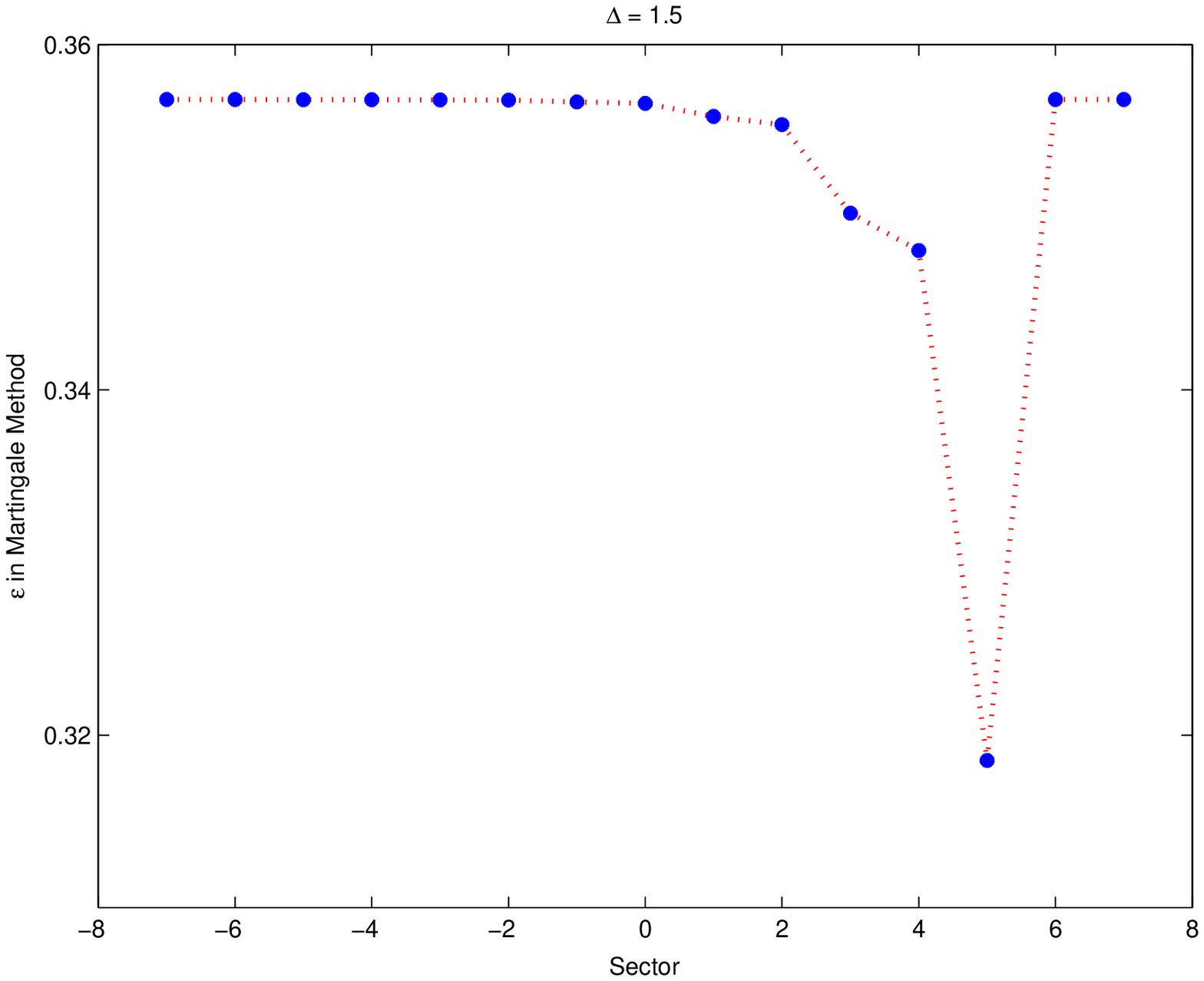}\ \includegraphics[scale=0.3]{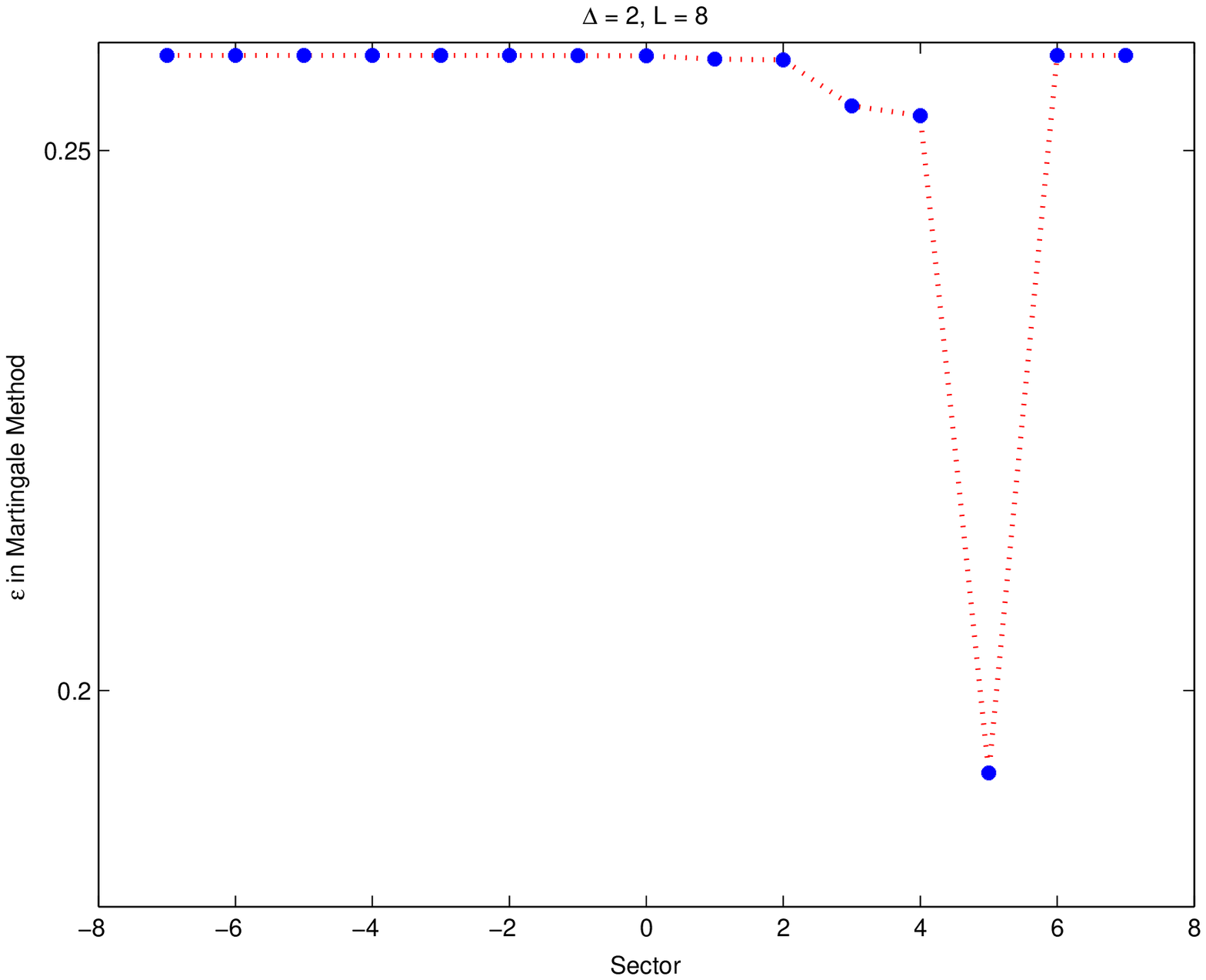}
\end{center}
\caption[$\varepsilon$ figures]{Some plots of $\varepsilon_m$ for a chain of length $L=8$ for different
values of $\Delta$.}
\end{figure}

The general shape persists for all values of
$L$ that we have tried (up to 11,) and moreover, for fixed $\Delta$ the values
of $\epsilon_m$ stabilize very quickly relative to the number of downspins, i.e.  
in the sector corresponding to $k$ downspins for chains of length
six, seven, or eight the $\varepsilon_{L-k}$ varies no
more that one part in a thousand.  We include our code at the end of this paper
in Appendix \ref{appendix: matlab code}.

It is easy to compute that in the Ising limit ($\Delta\rightarrow\infty$), we
get $\varepsilon_m = 0$ for all $m$.  We hope to prove Conjecture
\ref{conjecture: epsilon} for at least most values of $\Delta$ by performing
perturbation around the Ising case.  This should be possible since for long
chains the ground states of the XXZ Hamiltonian can be thought of as
perturbations of the Ising ground states.  One also notices that $\varepsilon_m$
becomes very small, i.e. close to the value in the Ising case, for relatively
small values of $\Delta$ ($\Delta\approx{2}$).  This behavior gives us a hope of
being able to prove Conjecture \ref{conjecture: epsilon} for at least some
values of $\Delta$.

The following is a plot comparing the true gap of the XXZ Hamiltonian for a chain of length $L$ to the lower bound given by Theorem \ref{theorem: lower bound}.  The anisotropy parameter $\Delta$ varies between 1 and 20.

\begin{figure}[h]
\begin{center}
\includegraphics[scale=0.5]{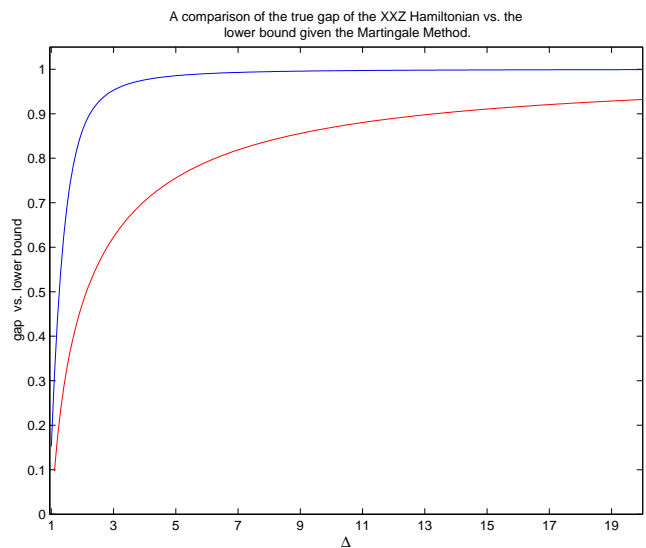}\
\end{center}
\caption[Comparison of true gap to lower bound]{The upper plot is the true spectral gap of the XXZ chain on $L$ sites.  The lower plot is the estimated gap from Theorem \ref{theorem: lower bound}.($1 \leq \Delta \leq 20$)}
\end{figure}

The plot shows that the estimate captures the behavior of the true gap very well.  Over the inteval $1 \leq \Delta \leq 20$, the estimate is only off by 17\% on average.  The maximum error occurs for $\Delta$ close to 1 and at this point the estimate is off by about 70\% from the true gap, but as $\Delta$ grows and gets closer to 20 the error shrinks to about 7\% of the actual gap.

\newpage
\pagestyle{myheadings} 
\markright{  \rm \normalsize CHAPTER 3. \hspace{0.5cm} 
  XY Model}
\chapter{XY Model}
\label{chapter: xy}
\thispagestyle{myheadings}

%%%%%%%%%%%%%%%%%%%%%%%%%%%%%%%%%%%%%%%%%%%%%%%%%%%%%%%%%%%%%%%%%%%%%%%%%%%%%%%%
%%%%%%%%%%%%%%%%%%%%%%% SECTION: INTRODUCTION %%%%%%%%%%%%%%%%%%%%%%%%%%%%%%%%%%
%%%%%%%%%%%%%%%%%%%%%%%%%%%%%%%%%%%%%%%%%%%%%%%%%%%%%%%%%%%%%%%%%%%%%%%%%%%%%%%%

\section{Introduction}

Understanding phase transitions, for example ice to water, or non-magnetic to 
magnetic, is one of the central questions of statistical mechanics.  Spin systems
were initially introduced by Lenz in 1920 to try to explain the tendency of some 
materials to possess macroscopic magnetic fields.  He proposed a simple model in which
atoms in a crystal could assume two possible orientations which later came to be
called the Ising model \cite{BR}.  It was Ising's failure, in 1925, to discover a phase
transition in Lenz's model which led Heisenberg to introduce quantum spin models
in 1928.  In 1935, Peierls gave an argument that demonstrated that the Ising
model does exhibit phase transitions in two and greater dimensions \cite{Pei}.  Although
Peierls' argument was not completely rigorous it did contain all the essential
ideas required for the desired result.  The Peierls' argument was later made 
rigorous by Griffiths \cite{Gri}, and Dobrushin \cite{Dob1}.

Mathematically, phase transitions are associated with a singularity in some
thermodynamic potential, generally the free energy.  While the Ising model is a 
classical model, it has become the basis for understanding magnetic phase 
transitions in quantum mechanical models.  The Ising Hamiltonian is given by
\be
\label{Ising Hamiltonian}
H^{I}_\Lambda(\{s_x\}) = -\sum_{\stackrel{x,y\in\Lambda}{|x-y|=1}}s_xs_y - h\sum_{x\in\Lambda}s_x
\ee
where $\Lambda$ is a finite subset of $\Ir^d$, $h$ is some positive number, and
$\{s_x\}$ is in the set $\Omega_\Lambda=\{\pm 1\}^\Lambda$.  The
partition function of the system $Z^I_\Lambda(h,\beta)$ is given by
\be
\label{Ising partition function}
Z^I_\Lambda(h,\beta) = \sum_{\{s_x\}\in\Omega_\Lambda}H^I_\Lambda(\{s_x\})
\ee
For any thermodyamic variable, $X$, defined on $\Omega_\Lambda$, we define the
expectation of $X$ at inverse temperature $\beta$ by 
\be
\label{Ising expectation}
\omega_{h,\beta}(X) =
\lim_{\Lambda\nearrow\Ir^d}\frac{1}{Z^I_\Lambda(h,\beta)}\sum_{\{s_x\}\in\Omega_\Lambda}X(\{s_x\})e^{-\beta H^I_\Lambda(\{s_x\})}
\ee
In dimension two and greater the Ising Model exhibits spontaeous magnetizaton.  
Consider the average magnetization at a particular site, 
$m(h,\beta) = \omega_{h,\beta}(s_0)$. Then there is a constant 
$\beta_c=\beta_c(d)$, such that for all $\beta>\beta_c$, $m(h,\beta)$ is 
discontinuous at $h=0$.  Specifically
\be
\lim_{h\rightarrow 0^-}m(h,\beta) = - \lim_{h\rightarrow
0^+}m(h,\beta)
\ee
for all $\beta>\beta_c(d)$.  This implies that the free energy, $f^I(h,\beta)$,
 is not analytic at $h=0$ since
\be
m(h,\beta) = \frac{\partial}{\partial h}\left(f^I(h,\beta)\right)\Big{|}_\beta
\ee
Physically, the discontinuity in $m(h,\beta)$ at $h=0$, can be interpreted as
the system ``remembering'' the effects of an external magnetic field after the
field has been removed, i.e. it continues to possess a macroscopic magnetic
field.

The phase diagram of the XY model has been the subject of much study. For high 
temperature, one can use a high-temperature expansion to show that there is a
unique phase.  To see an excellent discussion of such techniques see \cite{Rue}.
In \cite{DLS} Dyson, Lieb, and Simon showed that, in the absence of an external 
magnetic field, the spin-1/2 XY model does have a phase transition for positive 
temperature in dimensions greater than two.  The Mermin-Wagner \cite{BR} theorem 
shows that  for positive temperature this result can not be extended to 
dimension two. In a later paper \cite{KLS2}, Kennedy, Lieb, and Shastry 
demonstrated that in dimensions two and greater the ground state of the XY model 
has off-diagonal long-range order.  

\begin{figure}[h]
\begin{center}
\includegraphics[scale=0.7]{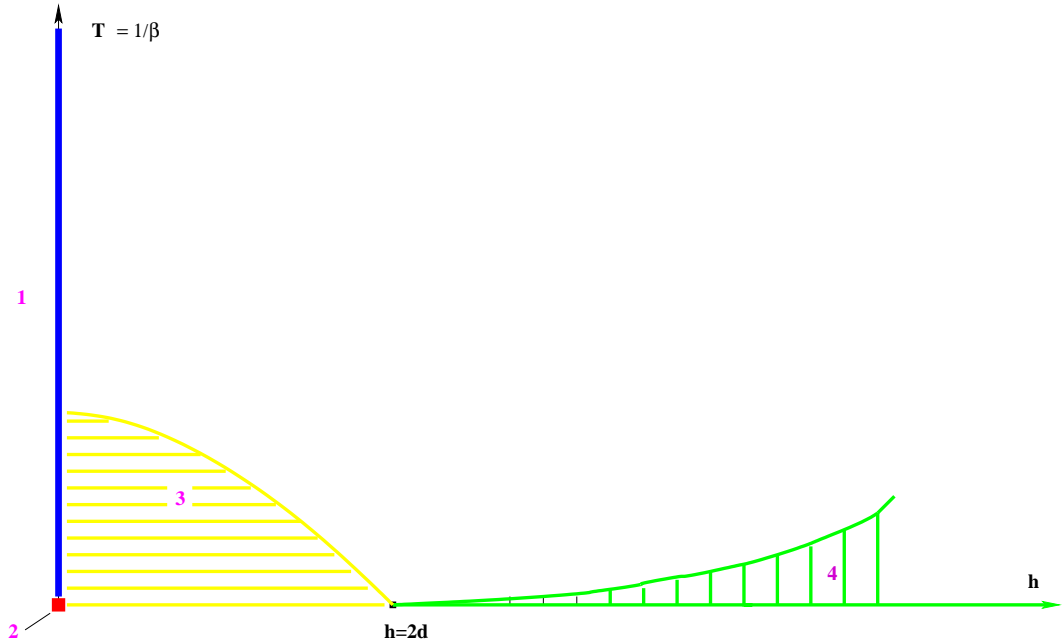}
\end{center}
\caption[Phase Diagram of the XY Model]{The phase diagram for the XY Model.  
The numbered regions are: {\bf{1}}. ($h=0$) For dimension three and greater 
Dyson, Lieb, and Simon showed that LRO exists for $T>0$. {\bf{2}}. For spins 
bigger than one-half and dimensions two or greater, Kennedy, Lieb, and Shastry showed the
groundstate has LRO. {\bf{3}}.  The low-temperature/small $h$ regime is the most 
interesting, but nothing is known (the shape of the curve is unknown.) 
{\bf{4}}.  We show that there is a low-temperature/large $h$ region in which there
is no phase transition.}
\end{figure}

Here we show that the free energy of the XY model in the 
presence of an external magnetic field above a critical strength, is analytic for 
some region of non-zero temperature in dimension two and greater.
The more interesting regime is when $h<h_c=2d$, where there is a connection to
Bose-Einstein condensation \cite{KLS2}. Our results do not extend to this case.

We consider a perturbation of the spin-$\frac{1}{2}$ XY model by a uniform 
magnetic field in the z direction.  The strength of the magnetic field is great 
enough so that there is a unique ground state of all ``up'' spins.  We use a 
contour expansion inspired by Kennedy in \cite{Ken} to extend the uniqueness of phase to 
non-zero temperatures.

Take $\HH_\Lambda=\Big(\Cx^2\Big)^{\bigotimes{|\Lambda|}}$ where 
$\Lambda\subset\Ir^d$, $d\geq{2}$, and $|\Lambda|<\infty$.  For $x,y\in\Lambda$ 
such that $|x-y|=1$ we write $\{x,y\}\equiv\nn{x}{y}$.  The Hamiltonian is 
defined as follows:
\begin{equation}
\label{Hamiltonian definition}
H'_\Lambda   = H^{XY}_\Lambda+hH^Z_\Lambda,\ \ \ \ h\geq{h}_c(d)=2d
\ee
where
\be
\label{XY part}
H^{XY}_\Lambda  =-\sum_{\nn{x}{y}\subset\Lambda}\frac{1}{2}(\s_x^1\s_y^1+\s_x^2\s_y^2)=
\sum_{\nn{x}{y}\subset\Lambda}(\s^+_x\s^-_y+\s^-_x\s^+_y)\\
\ee
and
\be
\label{magnetic part}
H^Z_\Lambda = \sum_{x\in\Lambda}\frac{1}{2}(\unity-\s_x^3)\\
\end{equation} 
The operators $\s^1,\ \s^2,$ and $\s^3$ are the Pauli spin matrices  
\begin{equation*}
\hspace*{\fill}\s^1=\left(\begin{array}{cc}
0&1\\
1&0
\end{array}\right);\ \ \ \ \s^2=\left(\begin{array}{cc}
0&-i\\
i&0
\end{array}\right);\ \ \ \ \s^3=\left(\begin{array}{cc}
1&0\\
0&-1
\end{array}\right)\hspace*{\fill}
\end{equation*}
where the subscript denotes the site on which they act.
We take the strength of the magnetic field $h$ to be greater that some $h_c(d)$ 
so that the unique state is the all up state $\ket{\uparrow}^\Lambda$ given by
\be
\label{all up state}
\ket{\uparrow}^\Lambda = \bigotimes_{x\in\Lambda}\ket{\uparrow}_x
\ee
For the remainder of this chapter we will work with the scaled
Hamiltonian
\begin{equation}
\label{working hamiltonian}
H_\Lambda \equiv (1-\delta)H^{XY}+h_cH^Z
\end{equation} 
where the parameter $\delta$ is given by the equation
\be
\label{delta definition}
1-\delta = \frac{h}{h_c}
\ee
By (\ref{Hamiltonian definition}), $h>h_c$ so we have $\delta\in(0,1)$.

Let 
\be
\label{partition function}
\mathrm{Z_\Lambda}(\delta,\beta)=\tr\ {e}^{-\beta{H_\Lambda}}
\ee
be the partition function.  Our goal is to show that the free energy
\be
\label{free energy}
f(\delta,\beta)=-\frac{1}{\beta}\ \lim_{\Lambda\nearrow\Ir^d}\ \
\frac{1}{|\Lambda|}\log\left[\tr\ {e}^{-\beta{H_\Lambda}}\right]
\ee
is analytic for all $(\delta,\beta)$ in some appropriate region, and hence that
no first-order phase transition exists.  In order to do this, we will express
$Z_\Lambda(\delta,\beta)$ as a contour expansion.  Then, we follow Kennedy in 
\cite{Ken}, and show that the partition function can be rewritten as  a cluster
expansion.  The key to this is Theorem \ref{proposition: key} which shows that
the weights of the contours decay in an appropriate way.  Once the cluster
expansion is established the analtyticity of $f(\delta,\beta)$ follows
immediately, as the cluster expansion is a uniform limit of analytic functions.
%%%%%%%%%%%%%%%%%%%%%%%%%%%%%%%%%%%%%%%%%%%%%%%%%%%%%%%%%%%%%%%%%%%%%%%%%%%%%%%%
%%%%%%%%%%%%%%%%%%%% SECTION: DEFINITIONS %%%%%%%%%%%%%%%%%%%%%%%%%%%%%%%%%%%%%%
%%%%%%%%%%%%%%%%%%%%%%%%%%%%%%%%%%%%%%%%%%%%%%%%%%%%%%%%%%%%%%%%%%%%%%%%%%%%%%%%

\section{Basic Definitions}

%Short appeal to origins of contour expansions.  Remark about Peierls argument
%Kennedy paper.

A set $\obs\subset\Lambda$ will be called a {\bf{contour}}. This differs from the 
usual definition of countours as set of bonds separating lattice sites with
opposite spins.  This is okay though, because in the model we consider the
ground state does not possess the spin-flip symmetry that generally introduces
ambiguities. A connected set $P\subset\Lambda$ will be called a {\bf{polymer}}.  
To each contour $\obs\subset\Lambda$ let $\ket{\obs}$ denote the basis vector of
$\HH_\Lambda$ such that
\be
\label{contour basis}
\ket{\obs} = \otimes_{x\in\Lambda}\ket{\obs}_x,\ \ \ket{\obs}_x =
\left\{\begin{array}{cc}
\ket{\obs}_x = \ket{\uparrow} & x\in\Lambda\setminus\obs\\
\ket{\obs}_x = \ket{\downarrow} & x\in\obs
\end{array}\right.
\ee
Furthermore, if $(\s^+_x\s^-_y+\s^-_x\s^+_y)\ket{\obs}\neq{0}$ then we will denote
the corresponding contour by $(\s^+_x\s^-_y+\s^-_x\s^+_y)\obs$.

Returning to $Z_\Lambda(\delta,\beta)$, by the Trotter product formula, we can express
$e^{-\beta{H}}$ as
$${e}^{-\beta{H_\Lambda}}=\lim_{N\longrightarrow\infty}\ \ \left[e^{-\frac{\beta{h}_c}{N}H^Z_\Lambda}\left(1-\frac{\beta(1-\delta)}{N}H^{XY}_\Lambda\right)\right]^N
$$
so that
\begin{eqnarray*}
Z_\Lambda(\delta,\beta) = \tr\ {e}^{-\beta{H_\Lambda}}&=&\lim_{N\longrightarrow\infty}\ \tr\ \left[e^{-\frac{\beta{h}_c}{N}H^Z_\Lambda}\left(1-\frac{\beta(1-\delta)}{N}H^{XY}_\Lambda\right)\right]^N \\
&=&\lim_{N\longrightarrow\infty}\ \sum_{\alpha\subset\Lambda}\bra{\alpha}\left[e^{-\frac{\beta{h}_c}{N}H^Z_\Lambda}\left(1-\frac{\beta(1-\delta)}{N}H^{XY}_\Lambda\right)\right]^N\ket{\alpha}
\end{eqnarray*}
inserting a resolution of the identity between each term in the product we can 
write the above as
\begin{eqnarray}
\tr\ {e}^{-\beta{H_\Lambda}}&=&\lim_{N\longrightarrow\infty}\ \sum_{\alpha_0,\ldots,\alpha_{N-1}}\bra{\alpha_0}\left[e^{-\frac{\beta{h}_c}{N}H^Z_\Lambda}\left(1-\frac{\beta(1-\delta)}{N}H^{XY}_\Lambda\right)\right]\ket{\alpha_1}\bra{\alpha_1}\ldots  \nonumber\\
&&\ldots\bra{\alpha_{N-1}}\left[e^{-\frac{\beta{h}_c}{N}H^Z_\Lambda}\left(1-\frac{\beta(1-\delta)}{N}H^{XY}_\Lambda\right)\right]\ket{\alpha_0} \label{eq:part} 
\end{eqnarray}
Taking this new expression for $Z_\Lambda(\delta,\beta)$ we see that the non-zero terms
in the sum are characterized by sequences of contours
$\{\alpha_0,\alpha_1,\ldots,\alpha_N\}$ with the following properties
\begin{enumerate}
\item[(i)]\ $\alpha_0=\alpha_N$
\item[(ii)]\ $\alpha_{i+1} = \left\{\begin{array}{cc}
\alpha_i&\\
\mathrm{or}&\\
(\s^+_x\s^-_y+\s^-_x\s^-_y)\alpha_i & \mathrm{for\ some}\ \nn{x}{y}
\end{array}\right.$
\end{enumerate}
This motivates the following definition

\begin{definition}
A {\bf{quantum contour}} $\Gamma=\{\Gamma_0,\Gamma_2,\ldots,\Gamma_N\}$ is a 
sequence of contours satisfying
\begin{enumerate}
\item[(i)]$\Gamma_0=\Gamma_N$
\item[(ii)]$\Gamma_{n+1}=\left\{
\begin{array}{cc}
\Gamma_n&\\
\mathrm{or}&\\
(\s^+_x\s^-_y+\s^-_x\s^+_y)\Gamma_n&\mathrm{where}\ |x-y|=1
\end{array}\right.$
\end{enumerate}
\end{definition}

\begin{figure}[h]
\begin{center}
\includegraphics[scale=0.5]{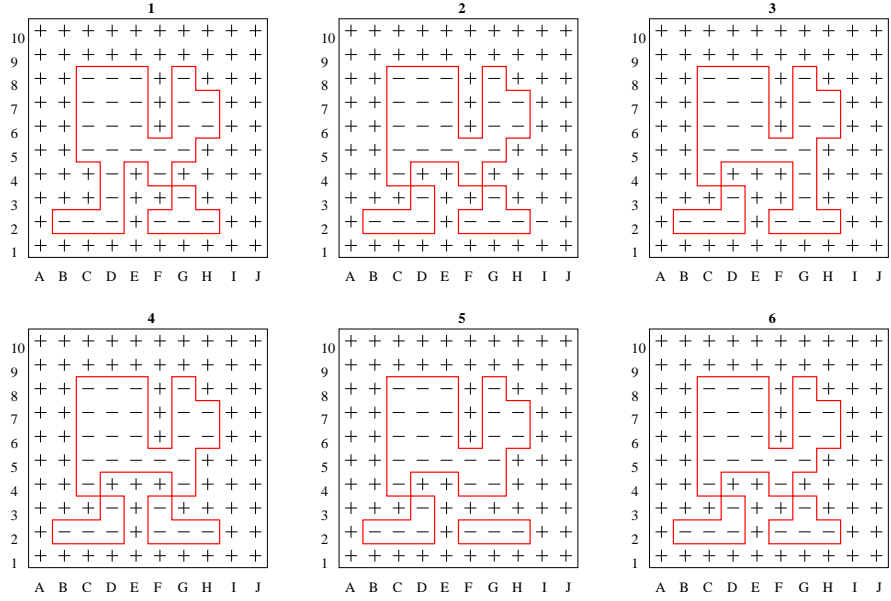}
\end{center}
\caption[Quantum Contour]{The spin-flips for the above quantum contour are
$1\rightarrow{2}:(C4,D4), 2\rightarrow{3}:(F4,G4), 3\rightarrow{4}:(F3,G3),
4\rightarrow{5}:(F3,F4), 5\rightarrow{6}:(G3,G4), 6\rightarrow{1}:(C4,D4)$}
\end{figure}

We now write
\bea
Z_\Lambda(\delta,\beta) &=& \lim_{N\longrightarrow\infty}\ \sum_{\alpha_0,\ldots,\alpha_{N-1}}\bra{\alpha_0}\left[e^{-\frac{\beta{h}_c}{N}H^Z_\Lambda}\left(1-\frac{\beta(1-\delta)}{N}H^{XY}_\Lambda\right)\right]\ket{\alpha_1}\bra{\alpha_1}\ldots  \nonumber\\
&& \ldots\bra{\alpha_{N-1}}\left[e^{-\frac{\beta{h}_c}{N}H^Z_\Lambda}\left(1-\frac{\beta(1-\delta)}{N}H^{XY}_\Lambda\right)\right]\ket{\alpha_0} \nonumber\\ 
&=& \lim_{N\rightarrow\infty} \sum_{\Gamma}\omega(\Gamma)
\eea
where
\be
\omega(\gamma) =
e^{-\beta{h_c}|\Gamma_0|}\left(\frac{\beta(1-\delta)}{N}\right)^{n(\Gamma)}
\ee
and $n(\Gamma)$ is the number of times $\Gamma_{i+1} =
(\s^+_x\s^-_y+\s^-_x\s^+_y)\Gamma_i$ in the quantum contour $\Gamma$. 

Let $S(\Gamma)={\bigcup}\Gamma_n$ be called the {\bf{support}} of a quantum 
contour $\Gamma$.   If $S(\Gamma)$ is connected, 
 then $\Gamma$ will be called {\bf{connected}}.We  will  use $\gamma$ to denote 
 a connected quantum contour.  Two connected quantum contours $\gamma^1$ and 
 $\gamma^2$ are disjoint if $S(\gamma^1)\cup{S}(\gamma^2)$ is not connected. 
 
\begin{figure}[h]
\begin{center}
\includegraphics[scale=0.5]{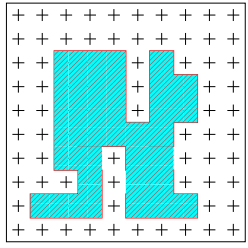}
\end{center}
\caption[Quantum Contour Support]{The support of the quantum contour above.  Note 
that it is connected.}
\end{figure}

 For a given quantum contour 
 $\Gamma$, $S(\Gamma)$ can be written as the disjoint union of connected parts 
$$S(\Gamma)=\bigsqcup_{j=1}^kS^j(\Gamma)$$
Then, $\Gamma$ can be split up into connected parts by
$$\gamma^j_n=\Gamma_n\cap{S}^j(\Gamma)$$
and we write $\Gamma=\sqcup\gamma^j$.  We call two general quantum contours
$\Gamma_1$ and $\Gamma_2$ disjoint if all there connected components are
disjoint in the sense of connected quantum contours defined above.

With these definitions fixed consider a quantum contour $\Gamma =
\Gamma^1\sqcup\Gamma^2$, then $\omega(\Gamma)$ will factorize in the following
way
$$
\omega(\Gamma)=e^{-\beta{h_c}|\Gamma_0|}\left(\frac{\beta(1-\delta)}{N}\right)^{n(\Gamma)}
$$
But, $\Gamma=\Gamma^1\sqcup\Gamma^2$ so that
\bea
|\Gamma_0| &=& |\Gamma^1_0| + |\Gamma^2_0|\\
n(\Gamma) &=& n(\Gamma^1) + n(\Gamma^2)
\eea
Thus
\bea
\omega(\Gamma) &=& e^{-\beta{h_c}|\Gamma^1_0|}\left(\frac{\beta(1-\delta)}{N}\right)^{n(\Gamma^1)}e^{-\beta{h_c}|\Gamma^2_0|}\left(\frac{\beta(1-\delta)}{N}\right)^{n(\Gamma^2)}\\
&=& \omega(\Gamma^1)\omega(\Gamma^2)
\eea
So that 
\begin{eqnarray}
\tr\ e^{-\beta{H_\Lambda}}\ &=&\ \lim_{N\longrightarrow\infty}\ \sum_\Gamma\omega(\Gamma) \nonumber\\
&=&\lim_{N\longrightarrow\infty}\sum_{n=0}^\infty\frac{1}{n!}\sum_{\stackrel{\gamma^1,\ldots,\gamma^n}{\mathrm{all\ disjoint}}}\prod_{j=1}^n\omega(\gamma^j)
\end{eqnarray}
Now, for a polymer $P$, we define
\be
\label{P polymer weights}
W(P) = \lim_{N\rightarrow\infty}\sum_{\gamma:S(\gamma)=P}\omega(\gamma)
\ee
Using these new weights we can re-write the partition function as
\bea
Z_\Lambda(\delta,\beta) &=&
\lim_{N\rightarrow\infty}\sum_{n=0}^\infty\frac{1}{n!}\sum_{\stackrel{\gamma^1,\ldots,\gamma^n}{\mathrm{all\ disjoint}}}\prod_{j=1}^n\omega(\gamma^j)\\
&=& \sum_{n=1}^\infty\sum_{\{P_1,\ldots,P_n\}}\prod_{i=1}^nW(P)
\eea
These are all the basic definitions which we need to apply the cluster expansion
%reference to earlier theorem here

%%%%%%%%%%%%%%%%%%%%%%%%%%%%%%%%%%%%%%%%%%%%%%%%%%%%%%%%%%%%%%%%%%%%%%%%%%%%%%%%
%%%%%%%%%%%%%%%%%%%%%% SECTION: CONTOUR EXPANSION %%%%%%%%%%%%%%%%%%%%%%%%%%%%%%
%%%%%%%%%%%%%%%%%%%%%%%%%%%%%%%%%%%%%%%%%%%%%%%%%%%%%%%%%%%%%%%%%%%%%%%%%%%%%%%%

\section{Cluster Expansion}
\label{sect:Cluster Expansion}

Cluster expansions were introduced as a perturbative method for high-temperature
gases.  They apply to continuous systems, e.g. classical interacting particle
systems, as well as discrete systems such as lattice particle models, polymer
models, or, as in our case, spin models.  There are many good references,
perhaps the most accessible being Simon's book on lattice gases \cite{Simon2}, 
while the treatise by Brydges \cite{Bry} is also excellent.
Others include \cite{GK},\cite{KP}, \cite{Pfi},\cite{Dob1}, \cite{BZ}, and \cite{Rue}
The basic goal of the cluster expansion is to give an expression for the
partition function of some system that can be more easily manipulated that the
standard expession as the sum of exponentials.  Here we present a condition for
the convergence of the cluster expansion attributed to Kotecky and Preiss \cite{KP}.  
The proof is taken from Ueltschi in \cite{Uel}. Although the setting of \cite{Uel}
is both continuous and discrete space, here we only need the discrete case.

Let $\mathbb{A}$ be a finite set whose elements are called {\bf{polymers}}.  Let
$\iota$ be a symmetric and reflexive relation on $\mathbb{A}\times\mathbb{A}$.
For polymers $A,A'\in\mathbb{A}$ we say that $A$ and $A'$ are
{\bf{incompatible}} if $A\iota{A'}$ and are {\bf{compatible}} otherwise.  The
partition function of the polymer model is
\be
Z(\mathbb{A}) = \sum_{n=1}^\infty\sum_{\{A_1,\ldots,A_n\}}\prod_{i=1}^nw(A_i)
\ee
where the $w$ is a complex function on $\mathbb{A}$ and the sum is over sets
of polymers whose elements are pairwise compatible.  

A sequence $C=(A_1,\ldots,A_n)$ of polymers is called a {\bf{cluster}} if
$\{A_1,\ldots,A_n\}$ can not be partitioned into two sets, all the elements in
the first being compatible with all the elements in the second.  We can make
this definition more visual.  Consider the graph $\mathcal{G}(C)$ with $n$
vertices where there is an edge between $i$ and $j$ whenever $A_i\iota{A_j}$.
$C$ is a cluster iff $\mathcal{G}(C)$ is connected.  We write 
$G\subset\mathcal{G}(C)$ if $G$ is a graph with the same number of vertices as
$\mathcal{G}(C)$ and whose edges form a subset of the edges of $\mathcal{G}(C)$;
$|G|$ denotes the number of edges of $G$.  Define a function, $\varphi(C)$, on 
clusters given by
\be
\label{cluster function 1}
\varphi(C) = \left\{\begin{array}{ll}
1& \mathrm{if}\ n=1\\
\sum_{G\subset\mathcal{G}(C),\mathrm{connected}}(-1)^{|G|}&\mathrm{if}\ n\geq{2}
\end{array}\right.
\ee
and a function $\Phi(C)$ on clusters by
\be
\label{cluster function 2}
\Phi(C) = \frac{1}{n!}\varphi(C)\prod_{i=1}^nw(A_i)
\ee
where $C=(A_1,\ldots,A_n)$.  The following theorem is due to Kotecky and Preiss \cite{KP}.
 It says that if the weight function on the polymers, $w$, satisfies some
 condition then the logarithm of the partition function $Z(\mathbb{A})$ can be
 given by a sum over clusters.
 
 %%%%%%%%%%%%%%%%%%%%%%%%%%%%% THEOREM: KP %%%%%%%%%%%%%%%%%%%%%%%%%%%%%%%%%%%%%
 
 \begin{theorem}
 \label{theorem: KP}
 Let $a$ and $b$ be non-negative functions on $\mathbb{A}$ such that for all
 $A\in\mathbb{A}$
 \be
 \label{KP-condition}
 \sum_{A',A'\iota{A}}|w(A')|e^{a(A')+b(A')} \leq {a}(A)
 \ee
 Then,
 \be
 \label{cluster expansion}
 Z(\mathbb{A}) = \exp\left(\sum_C\Phi(C)\right)
 \ee
 where the sum converges absolutely.  Furthermore
 \be
 \label{KP key}
 \sum_{C\iota{A}}|\Phi(C)|e^{b(C)} \leq a(A)
 \ee
 where $C\iota{A}$ means that $A$ is incompatible with at least one polymer in
 $C=(A_1,\ldots,A_n)$ and $b(C) = \sum_{i=1}^nb(A_i)$.
\end{theorem}

So, in order to show that $f(\delta,\beta)$ is analytic, we prove that
the weights of the polymers $W(P)$ given in (\ref{P polymer weights}) satisfies
condition (\ref{KP-condition}). Then Theorem \ref{theorem: KP} allows us to write
\bea
f(\delta,\beta) &=& -\frac{1}{\beta}\lim_{\Lambda\nearrow\Ir^d}
\frac{1}{|\Lambda|} \ln Z_\Lambda(\delta,\beta)\\
&=&-\frac{1}{\beta}\lim_{\Lambda\nearrow\Ir^d}\frac{1}{|\Lambda|}\sum_{\C}\Phi(\C)
\eea
which is the sum of analytic functions.  Then the analtycity of
$f(\delta,\beta)$ follows from the uniform convergence of this sum.  To this end
we state and prove the following proposition and leave the proof of Theorem
\ref{theorem: KP} to Appendix \ref{appendix: KP proof}.

%%%%%%%%%%%%%%%% PROPOSITION: KEY PROPOSITION %%%%%%%%%%%%%%%%%%%%%%%%%%%%%%%%%%
\begin{proposition}
\label{proposition: key}
Given $\delta$ where $0\leq\delta<1$, there exists a $\beta_0$ and an
$\varepsilon$ such that for all $\beta\geq\beta_0$
\be
\sum_{P\ni{y}}W(P)e^{\varepsilon|P|}\leq{\varepsilon}
\ee
where $y\in\Ir^2$ is fixed. 
\end{proposition}
Propostion (\ref{proposition: key}) immediately gives us (\ref{KP-condition})
since
\be
\sum_{P:P\cap{P}'\neq\emptyset}W(P)e^{\varepsilon|P|}\leq\sum_{y\in{P'}}\sum_{P\ni{y}}W(P)e^{\varepsilon|P|}\leq\sum_{y\in{P'}}\varepsilon\leq\varepsilon|P'|
\ee

\begin{proof}
Initially we will express the sum $\sum_{P\ni{y}}W(P)e^{\varepsilon|P|}$ as a
sum over contours.  Recall that 
$$W(P) = \lim_{N\rightarrow\infty}\sum_{\gamma:S(\gamma)=P}\omega(\gamma)$$
So, $\sum_{P\ni{y}}W(P)e^{\varepsilon|P|}$ can be re-written as 
\bea
\sum_{P\ni{y}}W(P)e^{\varepsilon|P|} &=&
\sum_{P\ni{y}}\left(\lim_{N\rightarrow\infty}\sum_{\gamma:S(\gamma)=P}\omega(\gamma)\right)e^{\varepsilon|P|}\\
&=&
\lim_{N\rightarrow\infty}\sum_{\gamma:y}\omega(\gamma)e^{\varepsilon|S(\gamma)|}\label{penalized weights}
\eea
where we have written $\gamma:y$ for the condition $y\in{S}(\gamma)$. 
Let the new Hamiltonian $H^0_\Lambda$ be given by
\be
\label{H^0 definition}
H^0_\Lambda = (1-\delta/2)\left(H^{XY}_\Lambda+h_cH^Z_\Lambda\right)
\ee
and define the modified weights $\omega_0(\gamma)$ by
\be
\label{omega_0 definition}
\omega_0(\gamma) =
e^{-\beta(1-\delta/2)h_c|\gamma_0|}\left(\frac{(1-\delta/2)\beta}{N}\right)^{n(\gamma)}
\ee
which are just the weights of a quantum contour $\gamma$ when $H_\Lambda$ is replaced by
$H^0_\Lambda$ in the expansion of the partition function $Z_\Lambda(\delta,\beta)$
The weights $\omega(\gamma)$ and $\omega_0(\gamma)$ are related by
\be
\omega(\gamma)=m(\gamma)\omega_0(\gamma)
\ee
where the multiplyer $m(\gamma)$ is 
\be
\label{multiplyer definition}
m(\gamma) = e^{-\beta\frac{\delta}{2}h_c|\gamma_0|}\rho^{n(\gamma)},\ \ \ \rho =
\frac{1-\delta}{1-\delta/2}<1
\ee
Thus, the sum on the right side of equation (\ref{penalized weights}) can be expressed as
\be
\label{penalized weights in H^0}
\sum_{\gamma:y}\omega(\gamma)e^{\varepsilon|S(\gamma)|}
= \sum_{\gamma:y}\omega_0(\gamma)e^{-\beta\frac{\delta}{2}h_c|\gamma_0|}\rho^{n(\gamma)}e^{\varepsilon|S(\gamma)|}
\ee

Let $F(\gamma)$ be the set of spins that flip at least once in $\gamma$ then
$F(\gamma)$ satisfies the inequalities
\bea
|S(\gamma)| &\leq& |F(\gamma)| + |\gamma_0| \label{support bound}\\
&\mathrm{and}&\\
|F(\gamma)| &\leq& 2n(\gamma) \label{flip bound}
\eea
Inequality (\ref{support bound}) follows since $F(\gamma)$ contains all the spins that flip and any $\gamma_i$ contains
all the spins that never flip, and we get inequality (\ref{flip bound}) from
the fact that for every occurence of a spin flip in $\gamma$ two spins change,
but it might be possible for certain spins to change more than once in the
course of a quantum contour $\gamma$. 

From inequalities (\ref{support bound}) and (\ref{flip bound}) we get
$|S(\gamma)| \leq 2n(\gamma) + |\gamma_0|$, which in turn implies
$$e^{\varepsilon|S(\gamma)|} \leq
e^{2\varepsilon{n}(\gamma)}e^{\varepsilon|\gamma_0|}.$$
From the above inequality we can determine 
\be
e^{-\beta\frac{\delta}{4}h_c|\gamma_0|}\rho^{\frac{n(\gamma)}{2}}e^{\varepsilon|S(\gamma)|} \leq 
e^{(-\beta\frac{\delta}{4}h_c+\varepsilon)|\gamma_0|}\cdot\left(\rho^{1/2}e^{2\varepsilon}\right)^{n(\gamma)} \leq 1
\ee
for $\beta$ large enough, and $\varepsilon$ small enough.
Applying this estimate to (\ref{penalized weights in H^0}) we get
\be
\label{basic omega_0 extraction}
\sum_{\gamma:{y}}\omega(\gamma)e^{\varepsilon|S(\gamma)|} \leq
\sum_{\gamma:y}\omega_0(\gamma)\rho^{\frac{n(\gamma)}{2}}e^{-\beta\frac{\delta}{4}h_c|\gamma_0|}
\ee
We now re-write the sum on the right side as a sum over initial configurations
$\gamma_0$ and then a sum over quantum contours with that intial state.  This
give the expression 
\be
\label{initial configuration}
\sum_{\gamma:y}\omega_0(\gamma)\rho^{\frac{n(\gamma)}{2}}e^{-\beta\frac{\delta}{4}h_c|\gamma_0|} = \sum_{G\subset{\Lambda}}\sum_{\gamma:G,y}\omega_0(\gamma)\rho^{\frac{n(\gamma)}{2}}e^{-\beta\frac{\delta}{4}h_c|G|}
\ee
where we have used $\gamma:G$ to indicate the sum is over quantum contours
$\gamma$ such that $\gamma_0 = G$.

Let $D_y(G)$ be the number defined as follows:\ \ A set $K\subset\Lambda$ will
be called ${\mathbf{G,y}}${\bf{-connecting}} if the set $(K\cup{G}\cup\{y\})\subset\Lambda$ is
connected.  $D_y(G)$ is then defined as
\be
\label{definition:D_y(G)}
D_y(G) = \min \{|K|: K\ \mathrm{is}\ G,y-\mathrm{connecting}\}
\ee
In other words, $D_y(G)$ is the minimum number of sites required to make
$G\cup\{y\}$ a connected set.  Since $\gamma$ is a connected contour,
$F(\gamma)$ is $\gamma_0,y$-connecting, hence
\be
D_y(\gamma_0) \leq |F(\gamma)| \leq 2n(\gamma)
\ee
This allows one to bound the term $\rho^{n(\gamma)/2}$ by
$\rho^{D_y(\gamma_0)/4}$.  The advantage to this is that now the term does not depend
on the choice of quantum contour, but only on its initial configuration.
Therefore, in (\ref{initial configuration}), we can write
\be
\label{only initial}
\sum_{G\subset{\Lambda}}\sum_{\gamma:G,y}\omega_0(\gamma)\rho^{\frac{n(\gamma)}{2}}e^{-\beta\frac{\delta}{4}h_c|G|} \leq \sum_{G\subset\Lambda}\rho^{\frac{D_y(G)}{4}}e^{-\beta\frac{\delta}{4}h_c|G|}\sum_{\gamma:G}\omega_0(\gamma)
\ee
We can drop the condition that $\gamma$ be connected since adding more terms
only increases the sum. Thus, by (\ref{basic omega_0 extraction}),(\ref{initial
configuration}), and (\ref{only initial}), we have
\be
\label{random walk times H^0}
\lim_{N\rightarrow\infty} \sum_{\gamma:y}\omega(\gamma)e^{\varepsilon|S(\gamma)|}
\leq \lim_{N\rightarrow\infty} \sum_{G\subset\Lambda}\rho^{\frac{D_y(G)}{4}}e^{-\beta\frac{\delta}{4}h_c|G|}\sum_{\Gamma:G}\omega_0(\Gamma)
\ee
However, the only term on the right side of (\ref{random walk times H^0}) which
depends on the length of the quantum contour $\Gamma$ is 
$$\sum_{\Gamma:G}\omega_0(\Gamma)$$ thus we can pass the limit through the
initial sum to get 
\bea
\lim_{N\rightarrow\infty}\sum_{\Gamma:G}\omega_0(\Gamma) &=& \tr
\left(e^{-\beta{H}^0_\Lambda}\circ\mathbb{P}_G\right)\\
&=& e^{-\beta\ip{G}{H^0_\Lambda G}}
\eea
where $\mathbb{P}_G=\ket{G}\bra{G}$.  But $H^0_\Lambda\geq{0}$, so
$e^{-\beta\ip{G}{H^0_\Lambda G}}\leq{1}$
Thus we have shown 
\bea
\lim_{N\rightarrow\infty} \sum_{\gamma:y}\omega(\gamma)e^{\varepsilon|S(\gamma)|}
&\leq&
\sum_{G\subset\Lambda}\rho^{\frac{D_y(G)}{4}}e^{-\beta\frac{\delta}{4}h_c|G|}\lim_{N\rightarrow\infty}\sum_{\Gamma:G}\omega_0(\Gamma)\nonumber\\
&\leq& \sum_{G\subset\Lambda}\rho^{\frac{D_y(G)}{4}}e^{-\beta\frac{\delta}{4}h_c|G|}
\eea
The proof is then completed by the following lemma.

%%%%%%%%%%%%%%%%%%%%%%%%%% LEMMA: LEMMA:R %%%%%%%%%%%%%%%%%%%%%%%%%%%%%%%%%%%%%%

\begin{lemma}
\label{lemma:r}
For fixed $\delta$, $0<\delta\leq{1}$
\be
\sum_{G\subset\Lambda}\rho^{\frac{D_y(G)}{4}}e^{-\beta\frac{\delta}{4}h_c|G|}
\leq r(\delta,\beta)
\ee
where $r(\delta,\beta)\longrightarrow{0}$ as $\beta\longrightarrow\infty$
\end{lemma}

To prove Lemma \ref{lemma:r} we follow Kennedy and think of $G$ as being
generated by a random walk on points on the lattice in the following way:
Let $F$ be a $G,y$-connecting set with $|F| = D_y(G)$.  Consider a 
nearest-neighbor walk $\bar{\eta}$ on the elements of $F\cup{G}\cup\{y\}$ where 
$\bar{\eta}_0=y$.  We take $\bar{\eta}$ in such a way that it does not visit any
site in $F\cup{G}\cup\{y\}$ more $c_0$ times, where $c_0$ is a constant that
only depends on the dimension.  We take $G$ to be generated by a walk $\eta$
where $\eta$ is the walk $\bar{\eta}$ with the steps on the set $F$ removed.
Notice that in general $\eta$ is not a nearest-neighbor walk. The product
\be
\rho^{\frac{D_y(G)}{4}}e^{-\beta\frac{\delta}{4}h_c|G|}
\ee
contains a factor $\rho^{1/4}$ for every site in $F$ and a factor
$e^{-\frac{\beta\delta{h}_c}{4}}$ for each site in G.  Taking
\be
\rho_0=\rho^{\frac{1}{4c_0}}\ \ \mathrm{and}\ \ \delta_0 =
\frac{\delta{h}_c}{4c_0}
\ee
gives
\be
\rho^{\frac{D_y(G)}{4}}e^{-\beta\frac{\delta}{4}h_c|G|} \leq
e^{-\beta\delta_0|\eta|}\prod_{i=1}^{|\eta|}\rho_0^{d(\eta_{i-1},\eta_i)}
\ee
where $d(\eta_{i-1},\eta_i)$ is the minimal number of sites required to connect
$\eta_{i-1}$ and $\eta_i$.  Therefore
\bea
\sum_{G\subset\Lambda}\rho^{\frac{D_y(G)}{4}}e^{-\beta\frac{\delta}{4}h_c|G|}
&\leq&
\sum_\eta
e^{-\beta\delta_0|\eta|}\prod_{i=1}^{|\eta|}\rho_0^{d(\eta_{i-1},\eta_i)}\\
&=&\sum_{n=1}^\infty
e^{-\beta\delta_0|\eta|}\sum_{\eta:|\eta|=n}\prod_{i=1}^n\rho_0^{d(\eta_{i-1},\eta_i)}
\eea
But
\bea
\sum_{\eta:|\eta|=n}\prod_{i=1}^n\rho_0^{d(\eta_{i-1},\eta_i)} &=&
\sum_{b_0,\ldots,b_n\in\Lambda}\prod_{i=0}^n\rho_0^{d(b_{i-1},b_i}\\
&\leq& \left(\sum_{b}\rho^{d(b_0,b)}\right)\cdots\left(\sum_{b}\rho^{d(b_n,b)}\right)
\eea
So, let $K=sup_{b'}\sum_{b}\rho_0^{d(b',b)}$.  Then
\be
\sum_{G\subset\Lambda}\rho^{\frac{D_y(G)}{4}}e^{-\beta\frac{\delta}{4}h_c|G|}
\leq \sum_{n=1}^\infty K^ne^{-\beta\delta_0n} = r(\delta,\beta)
\ee
and we note that $r(\delta,\beta)\longrightarrow{0}$ as
$\beta\longrightarrow\infty$.

This completes the proof of Lemma \ref{lemma:r} which in turn completes the
proof of Propostion \ref{proposition: key}.

\end{proof}

%%%%%%%%%%%%%%%%%%%%%%%%%%%%%%%%%%%%%%%%%%%%%%%%%%%%%%%%%%%%%%%%%%%
%%%%%%%%%%%%%%%%%%%%%%%%%%%%%%%%%%%%%%%%%%%%%%%%%%%%%%%%%%%%%%%%%%%

\appendix

\newpage
\pagestyle{myheadings} 
\markright{  \rm \normalsize APPENDIX A. \hspace{0.5cm} 
  Critical magnectic field $h_c(d)$}
\chapter{Calculation of critical magnetic field.}
\label{chapter: h_c}
\thispagestyle{myheadings}

In Chapter \ref{chapter: xy} above we presented a contour expansion for the free
energy of the system.  In this expansion the contours that contribute the most
are those that are close to the ground state of the Hamiltonian.  They can be
thought of as islands of down spins in a sea of up spins.  The argument relies
upon the fact that we only consider a region in which the all-up spin state is
the unique ground state of the system.
Here we present the proof that there is a critical magnetic field strength,
$h_c=2d$, such that for all $h>h_c$ the all spin up state is the unique ground
state of the XY Hamiltonian
$$
H^{XY} = -\sum_{|x-y|=y}(\s_x^+\s_y^-+\s_x^-\s_y^+) +
h\sum_{x}\frac{1}{2}(\idty-\s_x^3)
$$
We show $h_c \geq 2d$ by calculating the energy of spin waves and comparing it
to the energy of of the all-up state.  The inequality $h_c \leq 2d$ is obtained
by writing the Hamiltonian as the sum over two-site interactions and then 
determining  a condition on $h$ such that the all up state minimizes every
two-site interaction.

\begin{proposition}
There is a critical value $h_c$ such that for $h\geq{h_c}$ the ground state 
of $H_\mathrm{I}$ is the all 'up' state, while for $h<h_c$ this is no longer 
the case.  Moreover $h_c=2d$
\end{proposition}
\begin{proof}
To find the critical value of $h$ we first calculate when a spin wave has lower
energy than the all up state.
Let $\psi^k=\sum_xa_x\s^-\ket{\uparrow}$ where $a_x=e^{ikx}/|\Lambda|^{1/2}$.  Then
\beann
\ip{\psi^k}{H_\mathrm{I}\psi^k}&=&\ip{\psi^k}{H^{XX}\psi^k}+\ip{\psi^k}{H^Z\psi^k}\\
&=&-\sum_{\stackrel{\nn{x}{y}}{z,z'}}\bar{a}_za_{z'}\bra{\uparrow}\s_z^+(\s_x^+\s_y^-+s_x^-\s_y^+)\s_{z'}^-\ket{\uparrow}+\frac{h}{2}\sum_{\stackrel{x}{z,z'}}\bar{a}_za_{z'}\bra{\uparrow}\s_z^+(\unity-\s_x^3)\s_{z'}^-\ket{\uparrow}
\eeann
But $\bra{\uparrow}\s^+_z\s^+_x\s^-_y\s^-_{z'}\ket{\uparrow} =
\delta_{x,z'}\delta_{y,z}$ and $\bra{\uparrow}\s^+_z\s^-_x\s^+_y\s^-_{z'}\ket{\uparrow} =
\delta_{x,z}\delta_{y,z'}$, so
\bea
\ip{\psi^k}{H\psi^k} &=& -\sum_{\nn{x}{y}}\bar{a}_ya_x -
\sum_{\nn{x}{y}}\bar{a}_xa_y + h\\
&=& -2\sum_{\nn{x}{y}}\bar{a}_xa_y +h\\
&=&
-\frac{2}{|\Lambda|}\sum_{x\in\Lambda}e^{-ikx}\frac{1}{2}\sum_{j=1}^d(e^{ik(x+e_j)}+e^{ik(x-e_j)})  + h\\
&=& -2\sum_{j=1}^d\cos{k_j} + h \geq -2d + h
\eea
Therefore the spin wave $\psi^k$ at least has variational energy $-2d + h$.  But
the state $\ket{\uparrow}$ has energy zero, so we are looking for the regime in
which
\be
-2d + h < 0 \iff h < 2d
\ee
This implies 
\be
h_c \geq 2d
\ee
To get the other inequality we write
\be
H = \sum_{\nn{x}{y}}\left(H^{(2)}_{x,y} + \frac{h}{2d}\idty\right),\ \
H^{(2)}_{x,y} = -(\s^+_x\s^-_y+\s^-_x\s^+_y) - \frac{h}{4d}(\s^3_x+\s^3_y)
\ee
If we denote by $E_0$ and $e_0$ the ground state energies of $H$ and
$H^{(2)}_{x,y}$ respectively, then we see
\be
E_0 \geq \sum_{\nn{x}{y}}(e_0 + \frac{h}{2d})
\ee
Then we can calculate for which values of $h$ the state $\ket{\uparrow}$
minimizes each $H^{(2)}_{x,y}$. For these values of $h$ 
$$
E_0 = \sum_{\nn{x}{y}}(e_0 + \frac{h}{2d})
$$
The eigenvalues of $H^{(2)}_{x,y}$ are $\pm{1}$ and $\pm{h}/2d$.  The eigenvalue
$-h/2d$ corresponds to the state $\ket{\uparrow\uparrow}$ at the sites $x$ and
$y$.  Hence, for $\ket{\uparrow}$ to minimize $H^{(2)}_{x,y}$ $h$ must satisfy
\be
-\frac{h}{2d} < -1 \iff h>2d
\ee
which gives the inequality 
\be
h_c \leq 2d
\ee
Therefore, we have $2d \leq h_c \leq 2d$ which implies $h_c=2d$.
\end{proof}

\newpage
\pagestyle{myheadings} 
\markright{  \rm \normalsize APPENDIX B. \hspace{0.5cm} 
  Convergence of Cluster Expansion}
\chapter{Proof of Theorem \ref{theorem: KP}}
\label{appendix: KP proof}
\thispagestyle{myheadings}

We restate Theorem \ref{theorem: KP} here and present a proof due to Ueltschi
\cite{Uel}.  The proof is very technical and is included here for
the sake of completeness.  Let $\mathbb{A}$ be a finite set whose elements are
called polymers.  Let $\iota$ be a symmetric and reflexive relation on
$\mathbb{A}\times\mathbb{A}$.  For polymers $A,A'\in\mathbb{A}$ we say that $A$
and $A'$ are incompatible if $A\iota A'$ and are compatible otherwise.  The
partition function of the polymer model is 
$$Z(\mathbb{A}) =
\sum_{n=1}^\infty\sum_{\{A_1,\ldots,A_n\}}\prod_{i=1}^nw(A_i)$$
where $w$ is a complex function on $\mathbb{A}$ and the sum is over sets of
polymers whose elements are pairwise compatible.

\begin{theorem}
 
 Let $a$ and $b$ be non-negative functions on $\mathbb{A}$ such that for all
 $A\in\mathbb{A}$
 \be
 \label{KP-condition2}
 \sum_{A',A'\iota{A}}|w(A')|e^{a(A')+b(A')} \leq {a}(A)
 \ee
 Then,
 \be
 \label{cluster expansion2}
 Z(\mathbb{A}) = \exp\left(\sum_C\Phi(C)\right)
 \ee
 where the sum converges absolutely.  Furthermore
 \be
 \label{KP key2}
 \sum_{C\iota{A}}|\Phi(C)|e^{b(C)} \leq a(A)
 \ee
 where $C\iota{A}$ means that $A$ is incompatible with at least one polymer in
 $C=(A_1,\ldots,A_n)$ and $b(C) = \sum_{i=1}^nb(A_i)$.
\end{theorem}

The sequence of the proof is out of order with the
statements of the theorem.  We first prove inequality (\ref{KP key2}) as this
follows from the assumption (\ref{KP-condition2}).  We then use (\ref{KP key2}) to
prove the convergence of the cluster expansion on the right hand side of
(\ref{cluster expansion2}) and finally show that the partition function
$Z(\mathbb{A})$ can indeed be written as the cluster expansion.

\begin{proof}
In order to prove inequality (\ref{KP key2}) we induct on number of elements in 
the cluster.  For $n=1$,
$C=(A_1)$ then by assumption (\ref{KP-condition2}) we have
$$
\sum_{C\iota{A},|C|=1}|\Phi(C)|e^{b(C)} \leq a(C)
$$
Assume that for all $k \leq n$
\be
\label{induction assumption}
\sum_{C\iota{A},|C| \leq k}|\Phi(C)|e^{b(C)} \leq a(C)
\ee
and consider the sum over clusters with $n+1$ elements or less.  To bound this,
first sum over polymers $A_1$ incompatible with $A$ and then over remaining
polymers.  Thus, we have
\begin{multline}
{\sum_{C\iota{A},|C| \leq n+1} |\Phi(C)|e^{b(C)} \leq 
\sum_{A_1\iota{A}}\sum_{j=2}^{n+1}\frac{1}{(j-1)!}\sum_{A_2,\ldots,A_j}} \\
 \prod_{\ell=1}^j|w(A_\ell)|e^{b(A_\ell)}\Big{|}\sum_{\stackrel{G\subset\mathcal{G}(A_1,\ldots,A_j)}{\mathrm{connected}}}(-1)^{|G|}\Big{|}
\end{multline}
For a given graph $G$, let $(G_1,\ldots,G_k)$ be a sequence of  connected
subgraphs of $G$.  Let each subgraph $G_i$ have vertice set $V_i$ where
$V_i\cap V_j=\emptyset$ if $i \neq j$.  Furthermore, assume that
$V_1\cup\cdots\cup V_k=\{2,\ldots,n\}$. The sequence $(G_1,\ldots,G_n)$
defines a disconnected subgraph $G'$ of $G$ obtained by removing all edges
eminating from vertex 1.  Hence,
\begin{multline}
\Big{|}\sum_{\stackrel{G\subset\mathcal{G}(A_1,\ldots,A_j)}{\mathrm{connected}}}(-1)^{|G|}\Big{|} \leq \sum_{k \geq 1}\frac{1}{k!}\Big{|}\sum_{(G_1,\ldots,G_k)}\prod_{i=1}^k (-1)^{|G_i|}\sum_{G_i'}(-1)^{|G_i'|}\Big{|} \\
= \sum_{k \geq 1}\frac{1}{k!}\Big{|}\sum_{(G_1,\ldots,G_k)}\prod_{i=1}^k (-1)^{|G_i|}\Big{|}
\end{multline}
since $\sum_{G_i'}(-1)^{|G_i'|}=-1$.  Next we expand the sum over sequnces
$(G_1,\ldots,G_k)$.  First sum over the respective number of vertices
$m_1,\ldots,m_k$ with the condition that $m_1+\cdots+m_k = j-1$, and then choose
connected graphs for each choice of $m_1,\ldots,m_k$.  We can bound the sum over
$A_2,\ldots,A_j$ by summing over clusters of length $k$ that are incompatible
with the polymer $A_1$ such that $|C_i|=m_i$.  This gives
\begin{multline}
\sum_{C\iota A,|C| \leq n+1} |\Phi(C)|e^{b(C)} \leq \sum_{A_1 \iota A}
w(A_1)e^{b(A_1}\sum_{j=2}^{n+1}\sum_{k \geq 0} \frac{1}{k!}\\
\sum_{(m_1,\ldots,m_k)\vdash{j-1}}\sum_{\stackrel{C_1,\ldots,C_k}{C_i\iota
A_1,|C_i|=m_i}}\prod_{i=1}^k|\Phi(C_i)|e^{b(C_i)}
\end{multline}
Relaxing the condition that $(m_1,\ldots,m_k)\vdash{j-1} \leq n$ to $m_i \leq n$
for all $i$ and then summing over $j$ gives
\be
\sum_{C\iota A,|C| \leq n+1} |\Phi(C)|e^{b(C)} \leq \sum_{A_1 \iota A}
w(A_1)e^{b(A_1}\sum_{k \geq 0} \frac{1}{k!}\left[\sum_{C \iota A,|C| \leq n}
|\Phi(C)|e^{b(c)}\right]^k
\ee
We now use the induction hypothesis (\ref{induction assumption}) to bound the
term in the brackets by $a(A_1)$, which in turn bounds the sum over $k$ by
$e^{a(A_1)}$.  But, under the assumptions of the theorem (\ref{KP-condition2})
\be
\sum_{A_1\iota A}|w(A_1)|e^{b(A_1)+a(A_1)} \leq a(A)
\ee
which give (\ref{KP key2}) by induction.

To get the convergence of the cluster expansion in (\ref{cluster
expansion2}) note that any cluster is incompatible with its first element, so
\be
\label{absolute convergence of clusters}
\sum_{C}|\Phi(C)| \leq \sum_{A\in\mathbb{A}}\sum_{C\iota A} |\Phi{C}| \leq
\infty
\ee
by (\ref{KP key2}).  Thus we have established the convergence of the cluster
expansion, what remains is to show 
$$
Z(\mathbb{A}) = \exp \left(\sum_C\Phi(C)\right)
$$
This will be done by expanding $Z(\mathbb{A})$ and correctly rearranging terms
so that the sum over clusters becomes apparent.
\bea
Z(\mathbb{A}) &=&
\sum_{\stackrel{\{A_1,\ldots,A_n\}}{\mathrm{compatible}}}\prod_{i=1}^\infty
w(A_i)\\
&=&
1+\sum_{n \geq 1} \frac{1}{n!}\sum_{(A_1,\ldots,A_n)}\prod_{i=1}^n
w(A_i)\prod_{j<i}(1-\chi[A_i\iota A_j])\\
&=& 1+\sum_{n \geq 1} \frac{1}{n!}\sum_{(A_1,\ldots,A_n)}\prod_{i=1}^n
w(A_i)\sum_{\stackrel{G\subset\mathcal{G}(A_1,\ldots,A_n)}{\mathrm{connected}}}(-1)^{|G|}
\eea
Again we sum over partitions of the vertices $\underbar{m}=(m_1,\ldots,m_k)\vdash n$, first summing over the size of the
partition $k$, considering sequences of graphs $(G_1,\ldots,G_k)$ corresponding
to the partition $\underbar{m}$.  This gives
\begin{equation}\begin{split}
 Z(\mathbb{A}) &= 1+\sum_{n \geq 1}\sum_{k \geq
 1}\frac{1}{k!}\sum_{(m_1,\ldots,m_k)\vdash n} \frac{1}{m_1!\cdots m_k!}\\ 
 & \quad\quad\quad\quad\quad\quad\shoveright{\prod_{i=1}^k\left[\sum_{A_1,\ldots,\A_{m_i}}\prod_{j=1}^{m_i}w(A_j)\sum_{\stackrel{G\subset\mathcal{G}({A_1,\ldots,A_{m_i}})}{\mathrm{connected}}}(-1)^{|G|}\right]}\\
 &= 1 + \sum_{n\geq 1}\sum_{k\geq 1}\frac{1}{k!}\sum_{(m_1,\ldots,m_k)\vdash
 n}\prod_{i=1}^n\sum_{C,|C|=m_i}\Phi(C)\\
 &= \exp\left(\sum_C \Phi(C)\right)
\end{split}\end{equation}
where the last line follows since by (\ref{absolute convergence of clusters}) we
have absolute convergence of the cluster expansion.
\end{proof}

\newpage
\pagestyle{myheadings} 
\markright{  \rm \normalsize APPENDIX C. \hspace{0.5cm} 
  Matlab Code}
\chapter{Matlab Code}
\label{appendix: matlab code}
\thispagestyle{myheadings}

We include the Matlab code with which we produced the images in Chapter \ref{chapter: xxz}.  There are two main programs spin1kink.m and spin1compare.m. 
In spin1kink.m we evaluate $\|G_{[L,L+1]}E_L\|$ for $L+1=8$ in each sector of $S^3_{\mathrm{tot}}$.  This is accomplished by building the XXZ Hamiltonian using the spin operators Sz, Sx, and Sy defined below.  The Matlab function {\bf{kron}}  computes the tensor product of its inputs.  The projections onto ground states are produced by calculating ground state vectors and then creating the operator $\ket{v}\bra{v}$ where $v$ is the given ground state.  Finally we compute the proper eigenvalues and plot them versus the $S^3_{\mathrm{tot}}$ sector.\\

{\bf{spin1kink.m}}
\begin{verbatim}
%spin1kink.m
% SPIN 1 KINK HAMILTONIAN
clear;
delta = 50; q = delta -sqrt(delta^2-1); 
A = sqrt(1-delta^(-2));
L = 8;

Sz=[1 0 0;0 0 0;0 0 -1];
Sx=[0 1 0;1 0 1;0 1 0]/sqrt(2);
Sy=[0 1 0;-1 0 1;0 -1 0]/(i*sqrt(2));

hdiag= kron(Sz,Sz);
hhop = kron(Sx,Sx) + kron(Sy,Sy);
hbdry= kron(Sz,eye(3)) - kron(eye(3),Sz);
HDIAG = sparse(3^L,3^L);
HHOP = sparse(3^L,3^L);
HBDRY = sparse(3^L,3^L);
Sztot = sparse(3^L,3^L);
for a=1:(L-1)
  HDIAG = HDIAG + kron(speye(3^(a-1)), kron(hdiag,speye(3^(L-1-a) ) ) ); 
  HHOP = HHOP + kron(speye(3^(a-1)), kron(hhop,speye(3^(L-1-a))));
  HBDRY =  HBDRY + kron(speye(3^(a-1)), kron(hbdry,speye(3^(L-1-a))));
  Sztot = Sztot + kron(speye(3^(a-1)),kron(Sz,speye(3^(L-a))));
end
Sztot = Sztot + kron(speye(3^(L-1)),Sz);

Hkink = - HDIAG - 1/delta * HHOP -  A*HBDRY;
H =  Hkink + (L-1) * speye(3^L);

% Projection G onto the last two sites, [L,L+1]
h2 = -hdiag-1/delta*hhop - A*hbdry + speye(3^2);
[v,d] = eigs(h2,5,'sm');
G=0;
for k=1:5
    G = G + v(:,k)*transpose(v(:,k));
end
G = kron(speye(3^(L-1)), G);

for downspins = 1:2*L-1
for m = downspins-1: downspins+1
Proj =[];Proj = speye(3^L); 
for n=0:(2*L)
 if ne(n,m),
   Proj = Proj*(Sztot - (L-n)*speye(3^L))/(n-m);
 end;
end;
Hred = transpose(Proj)*(H - 0.2*speye(3^L))*Proj;
[V,D] = eigs(Hred + 0.2*speye(3^L),1,'sm');
psi(:,m-downspins+2) =V;
end

Psi(:,1) = kron(psi(:,1),[0,0,1]'); 
Psi(:,2) = kron(psi(:,2),[0,1,0]'); 
Psi(:,3) = kron(psi(:,3),[1,0,0]'); 

A=[];
for i=1:3
    for j=1:3
        A(i,j)=transpose(Psi(:,i))*G*Psi(:,j);
    end
end
epsi = eigs(A); e(:,downspins+1) = epsi(2);
end
e(:,1) = sqrt(q^2/(1+q^2) *(1-q^(2*L))/(1-q^(2*L+2)));
for i=1:2*L-1
  p(i)=e(2*L-i);
end

s=[-L+1:L-1];
plot(s,p,'b:*');
\end{verbatim}

The program spin1compare.m is very straightforward.  It calls two other programs, Hmltn.m and SectrProj.m.  Hmltn.m produces the XXZ Hamiltonian of inputted spin on a chain whose length is passed to the program.  SectrProj.m produces a projection onto the desired sector of $S^3_{\mathrm{tot}}$.  In spin1compare.m we loop through values of $\Delta$ between 1 and a specified number DeltaMax.  For each value of $\Delta$ spin1compare.m calculate the spectral gap of the output from Hmltn.m by projecting into each sector of $S^3_{\mathrm{tot}}$ and calculating the two smallest eigenvalues.  In each sector, 0 is one of the eigenvalues and the other is the spectral gap.  spin1compare.m then plots the spectral gap for each $\Delta$ along with the estimate from Theorem \ref{theorem: lower bound}.\\

{\bf{spin1compare.m}}
\begin{verbatim}
% spin1compare.m
%This code produces a graph that compares the true gap of the
%XXZ Hamiltonian with the estimate in this Thesis: Theorem 2.4.5
%The variables are:
% L = Length of the chain  J = spin of the system considered
% DeltaMax = this is the max value of Delta (anisotropy parameter) 

clear;
L = 8;
J = 1;
DeltaMax = 20;


for i=10:(10*DeltaMax)
    Delta = (i/10);
    H = Hmltn(L,J,Delta);      %Sets H to be the XXZ Hamiltonian 
                               %on spin-J chain of length L with 
                               %anisotropy Delta
    for j = -L+1:L-1
        P = SectrProj(L,J,j);      %This sequence calculates the 
                                   %gap of H by projecting into 
                                   %different sectors and then
        smallH = transpose(P)*H*P; %finding the two smallest e-values.  
                                   %Since there is a unique
        v = eigs(smallH,2,'sr');   %ground state in every sector the 
                                   %first non-zero eigenvalue
        u(L+j) = v(2);             %is the gap in that sector.
        smallH=[]; P = [];
    end
    gap(i-9) = min(u);            %This asks now for the minimum 
                                  %gap over all sectors

    
    gamma = 2.5 - sqrt(2.25-(2/Delta^2)); %This is the gap of the 
                                          %two-site Hamiltonian given 
                                          %in 2.4.75
    q = Delta - sqrt(Delta^2-1);   %q solves equation 
                                   %2Delta = q + q^(-1) on 
                                   %interval (0,1)
    epsilon = sqrt(((1-q^(2*L))/(1-q^(2*L+2)))*(q^2/(1+q^2))); 
                                   %This is the value calculated 
                                   %in Lemma 2.4.7
    gapest(i-9) = gamma * (1-sqrt(2)*epsilon)^2;  
                                   %This is the lower bound 
                                   %provided in Theorem 2.4.4  
    Deltavec(i-9) = Delta;
    H=[];  %frees up memory
end

plot(Deltavec, gap, 'b')
hold on
plot(Deltavec, gapest, 'r')
\end{verbatim}

We also include the programs Hmltn.m and SectrProj.m.\\
{\bf{Hmltn.m}}
\begin{verbatim}
function Hmltn=Hmltn(L,J,Delta)

%produces the spin J XXZ-Hamiltonian

%A = 0.1; gives Delta = 1.005

DelInv = 1/Delta; %sqrt(1-A*A);
A = sqrt(1-Delta^(-2));

%Dimension of 1-site Hamiltonian
dim1 = 2*J+1;
%Pauli Spin Matrix in third (or Z) direction
S3 = sparse(1:dim1,1:dim1,J-(0:2*J),dim1,dim1);             
%Spin-raising operator
Splus = sparse(dim1,dim1);
for j=0:(2*J-1)
  Splus = Splus + sparse(2*J-j,2*J-j+1,sqrt((2*J-j)*(j+1)),dim1,dim1);
end
%Spin lowering operator
Sminus = transpose(Splus);

%SS = Isotropic nearest neighbor interaction S(x).S(x+1)
SS = kron(S3,S3) + (1/2)*(kron(Splus,Sminus) + kron(Sminus,Splus));  
%Ising = Ising Nearest neighbor interaction S3(x)*S3(x)
Ising = kron(S3,S3);
%h = nearest neighbor interaction 
h = -DelInv*SS - (1-DelInv)*Ising+J^2*speye(dim1^2);
%H = sum over all nearest neighbor pairs of h(x,x+1) = Hamiltonian
H = sparse(dim1^L,dim1^L);
for x=1:(L-1)
 H = H + kron(eye(dim1^(x-1)),kron(h,eye(dim1^(L-1-x))));
end
%Bdry = boundary term A*(S3(x+1)-S3(x))
Bdry = J*A*(kron(speye(dim1^(L-1)),S3) - kron(S3,speye(dim1^(L-1))));
H = H +Bdry;

Hmltn=H;


\end{verbatim}

{\bf{SectrProj.m}}
\begin{verbatim}
function SectrProj=SectrProj(L,J,sector)


%produces a projection matrix onto the the subspace of total S3 "sector"
% for a spin J chain of length L

downspins = J*L-sector;

%Dimension of 1-site Hamiltonian
dim1 = 2*J+1;
%Pauli Spin Matrix in third (or Z) direction
S3 = sparse(1:dim1,1:dim1,J-(0:2*J),dim1,dim1);

%S3tot = total third-component of spin operator
S3tot = sparse(dim1^L,dim1^L);
for x=1:L, 
 S3tot = S3tot + kron(speye(dim1^(x-1)),kron(S3,speye(dim1^(L-x))));
end
%We now define the projection to the sector specified by
% S3tot = (J*L-downspins)
Proj = speye(dim1^L);
for n=0:(2*J*L)
 if ne(n,downspins),
   Proj = Proj*(S3tot - (J*L-n)*speye(dim1^L))/(n-downspins);
 end;
end;


%Proj is the orthogonal projection onto sector with specified number
% of downspins. Next we want to define a projection from this subspace
% to a vector space of the same dimension.
%The command "find" finds the nonzero elements of Proj, 
% which is what we need.
[I1,I2] = find(Proj);
dim = length(I1);
NewProj = sparse(I1,1:dim,ones(dim,1),dim1^L,dim);


SectrProj=NewProj;
\end{verbatim}

\newpage
\pagestyle{plain}
\addcontentsline{toc}{chapter}{{\bf Bibliography}}
\bibliographystyle{hplain}

\end{document}